\documentclass[11pt,a4paper]{article}

\pdfoutput=1
\usepackage{jheppub}
\usepackage[T1]{fontenc}
\usepackage[ansinew]{inputenc}
\usepackage[english]{babel}
\usepackage{color}
\usepackage{epsfig, palatino}
\usepackage{pstricks,pst-node,pst-tree}
\usepackage{epic}
\usepackage{mathrsfs}
\usepackage{ae} 
\usepackage{amsmath}
\usepackage{autobreak}
\usepackage{amssymb}
\usepackage{graphicx}
\usepackage{color}
\definecolor{darkblue}{cmyk}{0.9,0.9,0,0}
\usepackage{wasysym}
\usepackage{varioref}
\usepackage{makeidx}
\usepackage{simplewick}
\usepackage{array}
\usepackage{mathrsfs}
\usepackage{feynmf} 
\usepackage{tikz}
\usepackage{dsfont}
\usepackage{slashed}
\usepackage{float}
\usepackage{subcaption}
\usepackage{lscape}
\usepackage{afterpage}
\usepackage{soul}

\makeatletter
\DeclareRobustCommand*{\bfseries}{%
  \not@math@alphabet\bfseries\mathbf
  \fontseries\bfdefault\selectfont
  \boldmath
}

\makeatother

\usepackage[textsize=footnotesize]{todonotes}

\usetikzlibrary{decorations.pathmorphing}

\newcommand{\gammapsi}{\Psi}

\newcommand{\Fphiphibar}{F_{\phi\bar{\phi}}}
\newcommand{\Fphiphibarn}{F_{(\phi\bar{\phi})^n}}
\newcommand{\Fphiphibarfour}{F_{(\phi\bar{\phi})^2}}

\newcommand{\comment}[1]{}

\newcommand{\beq}{\begin{equation}}
\newcommand{\eeq}{\end{equation}}
\newcommand{\beqq}{\begin{equation*}}
\newcommand{\eeqq}{\end{equation*}}
\newcommand\beqa{\begin{eqnarray}}
\newcommand\eeqa{\end{eqnarray}}
\newcommand\beqaa{\begin{eqnarray*}}
\newcommand\eeqaa{\end{eqnarray*}}
\newcommand\bea{\begin{array}}
\newcommand\eea{\end{array}}

\newcommand{\nn}{\nonumber}

\newcommand{\neqa}{\nonumber\end{eqnarray}} 
\newcommand{\la}[1]{\label{#1}}

\renewcommand{\d}{\partial}

\newcommand{\<}{{\langle}}
\renewcommand{\>}{{\rangle}}

\newcommand{\cA}{{\cal A}}

\newcommand{\re}{\relax{\rm I\kern-.18em R}}

\renewcommand{\sp}{p\hspace{-.40em}/}

\newcommand{\ft}[2]{{\textstyle\frac{#1}{#2}}}

\newcommand{\phaneq}{\phantom{{}=}}

\def\XXint#1#2#3{{\setbox0=\hbox{$#1{#2#3}{\int}$}
\vcenter{\hbox{$#2#3$}}\kern-.5\wd0}}

\def\[{\left[}
\def\]{\right]}

\def\({\left(}
\def\){\right)}
\def\[{\left[}
\def\]{\right]}

\def\<{\langle}
\def\>{\rangle}

\def\i2{\frac{i}{2}}

\def\cO{{\mathcal O}}

\def\spi{\relax{\rm \pi\kern-0.5em /}}
\def\sA{\relax{\rm A\kern-0.5em /}}
\def\sp{\relax{\rm p\kern-0.5em /}}
\def\sd{\relax{\rm \d\kern-0.5em /}}
\def\sk{\relax{\rm k\kern-0.5em /}}
\def\sn{\relax{\rm n\kern-0.5em /}}
\def\sl{\relax{\rm l\kern-0.5em /}}
\def\sP{\relax{\rm P\kern-0.7em /}}
\def\sBethe{\relax{\rm \Bethe\kern-0.5em /}}

\def\cF{{\cal F}}

\def\One{1\hskip-.16cm1}

\def\cF{{\cal F}}

\def\cO{{\cal O}}

\def\cP{{\cal P}}

\def\cW{{\cal W}}

\def\2F1{\,_2{\rm F}_1}

\def \Ascr {\cA^{\,^{\kern-0.2em (\infty)}}}

\makeindex

\numberwithin{figure}{section}
\title{An Operator Product Expansion for Form Factors III.\\ Finite Coupling and Multi-Particle Contributions}

\author[a]{~~Amit Sever,}
\author[b]{~~Alexander G. Tumanov,}
\author[c]{~~Matthias Wilhelm}

\affiliation[a]{School of Physics and Astronomy, Tel Aviv University, Ramat Aviv 69978, Israel}
\affiliation[b]{Max-Planck-Institut f{\"u}r Physik, Werner-Heisenberg-Institut, F{\"o}hringer Ring 6, 80805 M{\"u}nchen, Germany}
\affiliation[c]{Niels Bohr Institute, University of Copenhagen,  Blegdamsvej 17, 2100 Copenhagen \O{}, Denmark}

\abstract{Form factors in planar $\mathcal{N}=4$ super-Yang-Mills theory have a dual description in terms of periodic Wilson loops. This duality maps the multi-collinear expansion of the former to an operator product expansion of the latter. The coefficients of this expansion are decomposed in terms of several elementary building blocks and can be determined at finite 't Hooft coupling using bootstrap and integrability techniques. Some of these building blocks are known from an analogous expansion of scattering amplitudes. 
In addition to these, the expansion for form factors includes a new type of building block, called {\it form factor transitions}, that encode information about the local operator.
In the present paper, we consider the form factor of the chiral part of the stress-tensor supermultiplet. We bootstrap the corresponding form factor transitions of two-particle flux-tube states and use them to predict the leading term in the collinear expansion at finite coupling. 
The transitions we find can be expressed  
in terms of a quantity that previously appeared in a seemingly unrelated context, namely the octagon kernel. Lastly, we use a factorized ansatz to 
determine the multi-particle form factor transitions at finite coupling, which we use to predict the first subleading term in the collinear expansion.
A perfect match is found between our predictions and the available perturbative data.
}

\preprint{MPP-2021-213}

\begin{document}
\maketitle

\newpage

\section{Introduction}
\label{sec: intro}

Form Factors (FFs) are fundamental quantities in Quantum Field Theory (QFT) that exhibit traits of both scattering amplitudes and correlation functions. They describe the amplitude of a state created by a local operator to decay into an $n$-point asymptotic state. In the maximally supersymmetric Yang-Mills theory in four dimensions ($\mathcal{N}=4$ SYM theory), they have been studied using modern on-shell methods (see, for instance, the review \cite{Yang:2019vag}) as well as integrability at weak coupling \cite{Frassek:2015rka} and strong coupling \cite{Maldacena:2010kp,Gao:2013dza}. 

In $\mathcal{N}=4$ SYM theory, planar $n$-point form factors have a dual description in terms of a certain type of periodic null polygonal Wilson loops $\cW_n$ with $n$ unique edges, which we also refer to as wrapped polygons \cite{Alday:2007he,Maldacena:2010kp,
Brandhuber:2010ad,Ben-Israel:2018ckc,Bianchi:2018rrj}. Based on this duality, in \cite{Sever:2020jjx,Sever:2021nsq} we have developed the {\it form factor operator product expansion} (FFOPE), used to compute FFs in planar ${\cal N}=4$ SYM theory non-perturbatively. In this approach, a generic wrapped
polygon is broken into a sequence of pentagon transitions $\cP$ \cite{Basso:2013vsa, Basso:2013aha,Basso:2014koa,Basso:2014nra,Basso:2014hfa,Basso:2015rta,Basso:2015uxa,Belitsky:2014sla,Belitsky:2014lta,Belitsky:2016vyq}. The sequence ends with a {\it form factor transition} $\cF$ that encodes the information about the local operator,
\beq\la{FFOPE}
\cW_n=\<\cF| e^{-H\tau_{n-2}+iP\sigma_{n-2}+iJ\phi_{n-2}}\cP\dots\cP e^{-H\tau_{1}+iP\sigma_{1}+iJ\phi_{1}}\cP|0\>\,.
\eeq
Here, $H,P,J$ are the Gubser-Klebanov-Polyakov (GKP) flux-tube Hamiltonian, momentum and angular momentum operators, respectively, and $\tau_i,\sigma_i,\phi_i$ are the cross-ratios that parametrize the polygon, with $\phi_{n-2}=0$; see \cite{Sever:2020jjx} for more details.

In this paper, we determine the FF transition of the chiral part of the stress-tensor supermultiplet at any value of the 't Hooft coupling. Together with the known spectrum of the GKP excitations \cite{Basso:2010in} and the pentagon transitions, this enables one to compute form factors at finite 't Hooft coupling as an expansion around the large-$\tau_i$ multi-collinear limit.

We determined the FF transition 
by solving a set of bootstrap axioms these objects satisfy \cite{Sever:2020jjx}. To uniquely fix the solution of the bootstrap axioms for the two-particle transitions, we use the Born-level two-particle transitions that we constructed in \cite{Sever:2021nsq} as well as a fruitful interplay with the perturbative bootstrap \cite{Dixon:2020bbt,PerturbativeBootstrap2}. Interestingly, we find that the FF transition for scalars is defined in terms of the so-called {\it octagon kernel}. 
This kernel has previously occurred in the study of the origin of the six-gluon amplitude \cite{Basso:2020xts}, as well as for the octagon form factor \cite{Coronado:2018ypq,Coronado:2018cxj,Kostov:2019stn,Kostov:2019auq,Belitsky:2019fan,Bargheer:2019exp}. The solutions for multi-particle GKP states are determined from the two-particle ones via a simple factorized ansatz, up to a coupling-independent matrix part that we explicitly construct for four particles.

This paper is structured as follows. In section \ref{sec: axioms}, we review the set of axioms that the FF transitions obey, as well as discuss some immediate consequences of them. Based on these axioms, in section \ref{sec: bootstrap} we bootstrap the finite-coupling expressions for the FF transitions of two particles. In section \ref{sec: factorization}, we bootstrap the FF transitions for more than two particles.
We end in section \ref{sec: conclusion} 
 with conclusions and an outlook. Six appendices include some  calculations needed to demonstrate the validity of our claims. 

\newpage

\section{Bootstrap axioms}
\label{sec: axioms}
\begin{figure}[t]
\centering
\includegraphics[width=7.5cm]{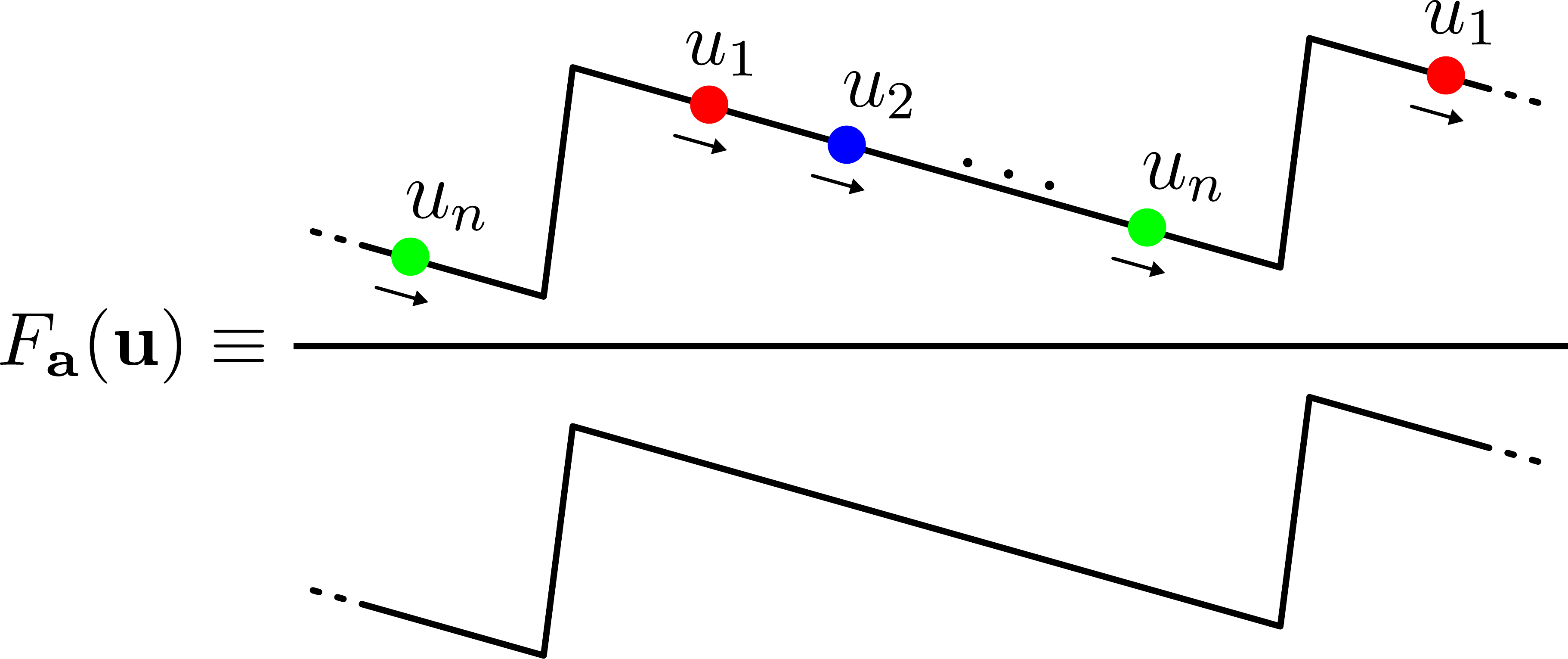}
\caption{The FF transition is defined as the amplitude of the two-sided wrapped polygon to absorb a GKP state. It is normalized in such a way that the FF transition for the GKP vacuum is equal to one. 
}
\label{FFtransdefinition}
\end{figure}

The FF transitions are subject to a set axioms that follow from their definition in figure~\ref{FFtransdefinition}, see \cite{Sever:2020jjx} for further details. To set the ground for the FF transition bootstrap, we summarize them here.

\begin{description}
\item[Watson] 
 Reordering two adjacent excitations within a GKP eigenstate is equivalent to acting on it with the S-matrix, which is known at any value of the 't Hooft coupling $\lambda$ \cite{Basso:2013pxa}. This property is inherited by the FF transition:
\begin{equation}\label{eq: Watson}
F_{\dots\,a_ja_{j+1}\,\dots}(\dots,u_j,u_{j+1},\dots)=\pm\, S(u_j,u_{j+1})_{a_ja_{j+1}}^{b_{j+1}b_j}F_{\dots\,b_{j+1}b_j\,\dots}(\dots,u_{j+1},u_{j},\dots)\, ,
\end{equation}
where the indices $b_j,b_{j+1}$ run over all flux-tube excitations, and the minus sign corresponds to the case where the two excitations are fermions, otherwise it is plus.
\item[Singlet] 
The two-sided wrapped polygon preserves the $U(1)_\phi\times SU(4)_R$ subgroup of the superconformal symmetry. Here, $U(1)_\phi$ parametrizes the rotations in the plane transverse to the polygon, while $SU(4)_R$ is the R-symmetry group. This implies that
\begin{equation}
\label{eq: singlet} 
F_{a_1\,\dots\,a_n}(u_1,\ldots,u_n)={\cal M}_{a_1}^{b_1}\dots{\cal M}_{a_n}^{b_n}\,F_{b_1\,\dots\,b_n}(u_1,\ldots,u_n)\, ,
\end{equation}
with ${\cal M}\in U(1)_\phi\times SU(4)_R$. As a result, FF transitions of charged states are equal to zero. 

The $U(1)_\phi\times SU(4)_R$ singlet states are easy to classify. All of the single-particle states are charged under at least one of the two symmetries. Therefore, the simplest singlet states are two-particle ones, consisting of two conjugate scalars $\phi\bar{\phi}$, two conjugate fermions $\psi\bar{\psi}$, or two conjugate gluon bound states $F_n F_{-n}$. All singlet states with more than two excitations are built from products of these elemental two-particle ones.

\item[Reflection]
The two-sided wrapped polygon is invariant under spacetime reflections. As a result, flipping the order of the GKP excitations and the signs of their momenta is a symmetry of the FF transition:
\begin{equation}
\label{eq: reflection}
F_{\bf a}({\bf u})=F_{\bar{\bf a}}(\bar{\bf u})\, .
\end{equation}
    Here, we use the abbreviation ${\bf a}=\{a_1,\dots,a_n\}$, ${\bf u}=\{u_1,\ldots,u_n\}$, $\bar{\bf a} = \{a_n,\ldots,a_1\}$ and $\bar{\textbf{u}}=\{-u_n,\dots,-u_1\}$.

\item[Crossing]
Applying two mirror transformations to the first excitation results in it being transferred to the last position:
\begin{equation}
\label{eq: crossing} 
F_{a_1a_2\dots a_n}(u_1^{2\gamma},u_2,\dots,u_n)=F_{a_2\dots a_na_1}(u_2,\dots,u_n,u_1)\,.
\end{equation}
Note that this transformation involves an analytic continuation which crosses certain cuts that shrink to zero in the perturbative regime. Its path depends on whether the $u_1$ excitation is a scalar, fermion or gluon; see \cite{Basso:2010in} for details.

\item[Square limit] 
The FF transition has a factorization pole at the kinematical point in which the momenta of the first and last excitations become equal,
\begin{align}\label{eq: squarelimit}
\lim_{u_1\to u_n}F_{\bf a}({\bf u})&=\frac{-\,i\delta_{a_n\bar a_1}}{\mu_{a_1}(u_1)}\frac{F_{a_2\,\dots\,a_{n-1}}(u_2,\dots,u_{n-1})}{ u_1-u_n-i\epsilon}\\
&\pm\frac{-\,i\delta_{b_n\bar b_1}}{\mu_{b_1}(u_1)}\frac{F_{b_2\,\dots\,b_{n-1}}(u_2,\dots,u_{n-1})}{ u_n-u_1-i\epsilon}\times\nn\\
&\qquad\quad\times\Big(\big[S_{12}\, S_{13}\ldots S_{1n-1}\big]\cdot S_{1n}\cdot\big[S_{n-1n}\dots S_{3n}\,S_{2n}\big]\Big)_{\bf a}{}^{\bf b}\, , \nn
\end{align}
where $S_{ij}\equiv S(u_i,u_j)$ and the square measure $\mu_a(u)$ is known at any value of the coupling \cite{Basso:2013vsa, Basso:2013aha,Basso:2014koa,Basso:2014nra}. Here, the plus sign corresponds to the case of a bosonic $u_1$ excitation and minus to the case of a fermionic $u_1$ excitation.  The first (second) term in this equation originates from the part of the GKP wave function in which the leftmost excitation is $u_1$ ($u_n$) and the rightmost excitation is $u_n$ ($u_1$). The two terms are related by the S-matrix prefactor, in accordance with the Watson axiom.

\end{description}

\section{Two-particle form factor transitions}
\label{sec: bootstrap}

In this section, we bootstrap the two-particle FF transitions at finite coupling using the axioms presented in the previous section as well as perturbative data.

\subsection{Gluons and Fermions}
\label{subsec: gluon and fermion transitions}

Let us consider the consequences of the square-limit axiom for the FF transitions of $\psi\bar\psi$ and $F_nF_{-n}$ first. 
Because the S-matrices for these excitations satisfy $S_{F_n F_{-m}}(u,u)=\delta_{nm}$ and $S_{\psi_A\bar\psi^B}(u,u)=-\,\delta_A^B$, the two terms in (\ref{eq: squarelimit}) combine into a delta function:
\beq\la{FpsiFFTsq}
\lim_{u\to v}F_{\psi_A\bar\psi^B}(u,v)=\frac{2\pi \delta_A^B}{\mu_{\psi}(u)}\,\delta(u-v)\,,\qquad
\lim_{u\to v}F_{F_nF_{-n}}(u,v)=\frac{2\pi}{\mu_{F_n}(u)}\,\delta(u-v)\,.
\eeq

In \cite{Sever:2021nsq} we have computed the transitions for $\psi\bar\psi$ and $F_1F_{-1}$ at Born level and found that they are given solely by their square-limit expressions. It is therefore natural to assume that the transitions for $\psi\bar\psi$ and $F_nF_{-n}$ are supported only on their square limit.  
In other words, these excitations cannot tell the difference between the wrapped two-sided polygon and the square:
\beq\la{FpsiFFT}
F_{\psi_A\bar\psi^B}(u,v)=\frac{2\pi \delta_A^B}{\mu_{\psi}(u)}\,\delta(u-v)\,,\qquad
F_{F_nF_{-n}}(u,v)=\frac{2\pi}{\mu_{F_n}(u)}\,\delta(u-v)\,.
\eeq

It is easy to see that the transitions in (\ref{FpsiFFT}) are consistent with the Watson and reflection axioms.\footnote{The measures are symmetric under reflection and crossing 
and are given in \cite{Basso:2013vsa, Basso:2013aha,Basso:2014koa,Basso:2014nra}.} The remaining crossing axiom is a little delicate to check. The FF transitions are distributions, that should be inserted under an integral over the rapidities. The crossing analytic continuation, $2\gamma$, should thus be understood as taking place inside an integral. To be concrete, we can realize a delta function as the difference between two poles, one above and one below the contour of integration. Its contribution to the integral is invariant under any analytic continuation of the path, and under $2\gamma$ in particular.

Another supporting argument for the solution 
(\ref{FpsiFFT}) comes from considering transitions with more than two excitations. 
One of the celebrated manifestations of integrability is the factorization of multi-particle fundamental quantities such as the S-matrix and the pentagon transitions into products of the two-particle ones. Likewise, we also expect the FF transitions to factorize. 
The square-limit axiom, however, contains a product of S-matrices in one of the two terms in (\ref{eq: squarelimit}), but not in the other. If this product is not equal to $\pm1$, then this fact would be inconsistent with a factorized solution. 
The solution (\ref{FpsiFFT}) forces the gluons and fermions into pairs of conjugate states with the same rapidity. As explained in section \ref{sec: factorization}, this results in these pairs dropping out from the product of S-matrices in the square limit of any other two excitations.\footnote{For the full consistency of a factorized ansatz, the scalars should also drop out of the product of S-matrices in (\ref{eq: squarelimit}). In section \ref{sec: factorization}, we show that this is indeed the case.}

Lastly, let us remark that the FF transition for the bound states $F_{n}$ and $F_{-n}$ follow from those of $F_{1},F_{-1}$ and the factorized ansatz for multi-particle transitions via so-called fusion, as we will elaborate on in section \ref{sec: factorization}.

\subsection{Scalars}\label{subsec: scalar transitions general}

Let us now turn to the scalars.
An immediate consequence of the singlet axiom \eqref{eq: singlet} is that only the singlet $SO(6)$ tensor structure is present\footnote{Note that in \cite{Sever:2020jjx,Sever:2021nsq} we do not write the flavor indices, as the sum over them is already explicitly performed in all the expressions.}
\beq\la{twoscalars}
F_{\phi^i\phi^j}(u,v)=\Fphiphibar(u,v)\times \delta^{ij}\,.
\eeq

Consider now the square-limit axiom. Because $S_{\phi\bar\phi}(u,u)=-1$, it takes the form
\beq\la{slscalar}
\lim_{u\to v}\Fphiphibar(u,v)=\frac{2i}{\mu_{\phi}(u)}\frac{1}{u-v}\,,
\eeq
where the $i\epsilon$ in the square-limit axiom leads to a principle-part  prescription for the pole.
Hence, unlike for gluons and fermions, the scalar transition cannot be localized at equal rapidities. Indeed, 
in \cite{Sever:2020jjx,Sever:2021nsq}, we found that at Born level it is given by 
\begin{equation}\label{eq: scalars at Born level}
\Fphiphibar(u,v)=-\,\frac{4}{g^2\,(u-v-2i)\,(u-v-i)}\,\frac{\Gamma\left(iu-iv\right)}{\Gamma\left(\frac{1}{2}+iu\right)\Gamma\left(\frac{1}{2}-iv\right)}+\mathcal{O}(g^0)\, ,
\end{equation}
where $g^2=\frac{\lambda}{16\pi^2}$.
In this subsection, we construct a minimal solution to the bootstrap axioms of $\Fphiphibar$, following similar steps to those taken  for the bootstrap of the scalar pentagon transition in \cite{Basso:2013aha}.

The Watson axiom \eqref{eq: Watson} for $\Fphiphibar$ 
takes the form
\begin{equation}
\label{eq: Watson special}
\Fphiphibar(u,v)=S_{\phi\bar\phi}(u,v)\Fphiphibar(v,u)\,,
\end{equation}
where $S_{\phi\bar\phi}$ is the S-matrix in the singlet channel, see \cite{Basso:2010in} and appendix \ref{S-matrixapp} for a review. This relation can be rewritten as
\begin{equation}
\frac{\Fphiphibar(u,v)^2}{\Fphiphibar(v,u)^2}=S_{\phi\bar\phi}^2(u,v)=\frac{S_{\phi\bar\phi}(u,v)}{S_{\phi\bar\phi}(v,u)}\,,
\end{equation}
where in the last step we used the unitarity of the S-matrix, $S_{\phi\bar\phi}(u,v)S_{\phi\bar\phi}(v,u)=1$. It can be solved by 
\begin{equation}
\Fphiphibar(u,v)^2=z(u,v)\,S_{\phi\bar\phi}(u,v)=\frac{(u-v+2i)(u-v+i)}{(u-v-2i)(u-v-i)}\,z(u,v)\,S_{\phi\phi}(u,v)\,,
\end{equation}
where $z(u,v)=z(v,u)$ is a symmetric function and in the last step we have used the relation \eqref{eq: singlet in terms of identical S-matrix}  between the scalar S-matrices in the symmetric channel ($S_{\phi\phi}$) and the singlet channel ($S_{\phi\bar\phi}$). 

Inserting this form of the FF transition into the crossing axiom \eqref{eq: crossing} yields a crossing relation for the symmetric function $z(u,v)$: 
\begin{equation}
\frac{z(u,v)}{z(u^{2\gamma},v)}=\frac{(u-v+4i)(u-v+3i)(u-v+2i)}{(u-v-2i)(u-v-i)(u-v)}\frac{S_{\phi\phi}(u^{2\gamma},v)}{S_{\phi\phi}(v,u)}\,, 
\end{equation}
where we used that for scalars the mirror analytic continuation is simply $u^\gamma=u+i$.
Next, we use crossing symmetry of the scalar S-matrix \cite{Basso:2013aha},
\begin{equation}
\label{eq: crossing of S matrix}
S_{\phi\phi}(u^{2\gamma},v)\,S_{\phi\phi}(u,v)=\frac{u-v}{u-v+2i}\,,
\end{equation}
and unitarity of the S-matrix to find
\begin{equation}
\frac{z(u,v)}{z(u^{2\gamma},v)}=\frac{(u-v+4i)(u-v+3i)}{(u-v-2i)(u-v-i)}\,.
\end{equation}
This can be solved via 
\begin{equation}\label{www}
z(u,v)=w(u,v)\times \frac{1}{(u-v-2i)(u-v-i)(u-v)(u-v+i)(u-v+2i)}\,,
\end{equation}
where
\beq\la{wcross}
w(u,v)=w(u^{2\gamma},v)\,,\qquad w(u,v)=-\,w(v,u)\,.
\eeq
In terms of the function $w(u,v)$, the square-limit axiom reads
\beq\la{wsquare}
\lim_{u\to v}w(u,v)=\frac{16}{\mu_{\phi}(u)^2}\frac{1}{u-v}\,.
\eeq
At leading order in the coupling, we find $w(u,v)$ by comparison with \eqref{eq: scalars at Born level}:
\begin{equation}
\label{eq: leading order w}
 w(u,v)=\frac{16}{\pi g^4}\,\frac{\cosh(\pi u)\cosh(\pi v)}{\sinh(\pi(u-v))}+\mathcal{O}(g^6)
 \,.
\end{equation}
Hence, a minimal finite-coupling solution for $w(u,v)$ that is consistent with (\ref{wcross}), (\ref{wsquare}), and (\ref{eq: leading order w}) is
\begin{equation}\label{eq: naive ansatz for the scalar w}
w_\text{min}(u,v)=\frac{16\pi}{\mu_\phi(u)\mu_\phi(v)}\frac{1}{\sinh(\pi(u-v))}\,.
\end{equation}
As we will see later, this ansatz \eqref{eq: naive ansatz for the scalar w} does not reproduce the next-to-leading order (NLO) data, though. We can parametrize the mismatch using a crossing-invariant symmetric Castillejo-Dalitz-Dyson (CDD) factor 
\begin{equation}
\label{eq: less naive ansatz for the scalar w}
w(u,v)=w_\text{min}(u,v)\times
G(u,v)^2\,.
\end{equation}
It is subject to the constraints
\beq
G(u,v)^2=G(v,u)^2\,,\qquad G(u,v)=1+\cO(g)\,,\qquad G(u^{2\gamma},v)^2 = G(u,v)^2\,.
\eeq
In terms of the CCD factor $G$, the FF transition reads
\beq\label{eq: scalar F in terms of G}
\Fphiphibar(u,v)^2=\frac{S_{\phi\bar\phi}(u,v)}{((u-v)^2+4)((u-v)^2+1)}\,\frac{16\pi}{\mu_\phi(u)\mu_\phi(v)}\,\frac{G(u,v)^2}{(u-v)\sinh(\pi(u-v))}\,,
\eeq
with the finite-coupling expressions for the measure and the S-matrix given in (\ref{measureFC}) and (\ref{SmatrixFC}), respectively.
Combining the constraints $G(u,v)^2=G(v,u)^2$ and $G(u,v)=1+\cO(g)$, we see that $G(u,v)=G(v,u)$. 
We will later find that $G(u^{2\gamma},v)=-\,G(u,v)$.\footnote{In contrast to the Watson relation, crossing does not commute with taking the weak-coupling limit.}

\subsection{Determining the scalar CDD factor with perturbative data}
\label{subsec: scalar transitions weak}

We now use perturbative data to find a finite-coupling expression for the CCD factor $G(u,v)=1+\mathcal{O}(g^2)$ that remained undetermined in the previous subsection. This perturbative data originates from the remainder function of the three-point form factor, which has been calculated up to two-loop order via unitarity \cite{Brandhuber:2012vm} and was subsequently bootstrapped up to eight-loop order \cite{Dixon:2020bbt,PerturbativeBootstrap2}.

While the two-loop unitarity calculation \cite{Brandhuber:2012vm} is naturally independent, the perturbative bootstrap \cite{Dixon:2020bbt,PerturbativeBootstrap2} made significant use of the FFOPE data, which in turn depends on the FF transitions and thus the function $G$ that is to be determined in this subsection. 
Before diving into the details of determining $G$, let us briefly explain this interplay between the FFOPE and the perturbative bootstrap.

The NLO correction to $G$ can be determined by requiring it to reproduce the $e^{-2\tau}\tau^0$ contribution to the large-$\tau$ expansion of the two-loop remainder. 
As elaborated on in \cite{Sever:2020jjx,Sever:2021nsq}, the operator product expansion of the three-point form factor consists of terms proportional to $e^{-\tau E}$, with $E$ being the energy of the corresponding GKP state, that is known at any value of the coupling \cite{Basso:2010in}. The loop corrections to $E$ are the only source of polynomial terms in $\tau$, which means that the term proportional to $e^{-2\tau}\tau^k$ is guaranteed to have at least $k$ powers of $g^2$. Because of this, the $\ell$-loop correction to the $e^{-2\tau}\tau^k$ term only receives contributions from the FF transition up to N${}^{\ell-k-1}$LO.
Thus, knowing $G$ at N${}^{k}$LO fully determines the $e^{-2\tau}\tau^{\ell-k-1}$ corrections to the form factor at $\ell$-loop order. This data can then be fed into the perturbative bootstrap to fully determine the remainder function at $(k+1)$-loop order, from which the N${}^{k+1}$LO correction to $G$ can be fixed. This sequence is repeated to obtain the next order correction, and so on.
This interplay between the FFOPE and the perturbative bootstrap is summarized in figure \ref{fig: interplay of perturbative and form factor transition bootstrap}.

\begin{figure}
\begin{align*}
G\text{ at N}^{k}\text{LO}\xrightarrow{\text{FFOPE}}\begin{array}{c}
e^{-2\tau}\tau^{\ell-k-1}\text{ term}\\
\text{at $\ell$ loops}
\end{array}\xrightarrow{\tiny\begin{array}{c}
\text{perturbative}\\
\text{bootstrap}
\end{array}}\begin{array}{c}
\text{$(k+1)$-loop}\\
\text{remainder}
\end{array}\longrightarrow G\text{ at N}^{k+1}\text{LO}
\end{align*}
\caption{Interplay between the FF transition bootstrap and the perturbative bootstrap.}
\label{fig: interplay of perturbative and form factor transition bootstrap}
\end{figure}

To determine the function $G(u,v)$, we first write an ansatz for it that is built 
from $\gammapsi^{(n)}(\frac{1}{2}\pm iu)$ and $\gammapsi^{(n)}(\frac{1}{2}\pm iv)$ as well as products thereof, where $\gammapsi^{(n)}(z)=\frac{d^{n+1}}{dz^{n+1}}\log \Gamma(z)$ is the polygamma function of order $n$. These functions are the building blocks occurring in the loop corrections to the pentagon transitions, the measures, the energy and the momentum of the scalar, and it is thus natural to expect them to occur also in the loop corrections to the FF transitions.
Moreover, we observe that when we assign weight $n+1$ to $\gammapsi^{(n)}$, only functions up to weight $m$ occur at order $g^m$ for quantities that start at order $g^0$. We thus assume this property in the ansatz, since $G(u,v)=1+\mathcal{O}(g^2)$. 
Next, we impose $G(u,v)=G(v,u)$ due to the Watson axiom, $G(u,u)=1$ due to the square-limit axiom and $G(u,v)=G(-v,-u)$ due to the reflection axiom. 
Note that we cannot impose crossing at the perturbative level, since the branch cut through which we are supposed to analytically continue closes at weak coupling. The remaining coefficients in this ansatz can be fixed by comparing the FFOPE prediction in terms of this ansatz with the data from the perturbative bootstrap. 
This way, we found
\begin{equation}
\label{eq: perturbative G}
\begin{aligned}
\log[&G(u,v)]=\frac{\pi^2g^2}{2}\left(t(u)-t(v)\right)^2\\
&+\frac{\pi^4g^4}{12}\left(t(u)-t(v)\right)^2 \left[15\,t(u)^2+15\,t(v)^2+18\,t(u)t(v)-16\right]\\
&+\frac{\pi^6g^6}{45}\left(t(u)-t(v)\right)^2 \left[165\,t(u)^4+165\,t(v)^4+240\,t(u)^3t(v)+240\,t(u)t(v)^3\right.\\
&\left.+\,270\,t(u)^2t(v)^2-300\,t(u)^2-300\,t(v)^2-360\,t(u)t(v)+136\right]+\mathcal{O}(g^8)\,.
\end{aligned}
\end{equation}
where we used the shorthand notation $t(u) = \tanh(\pi u)$.
Note that the logarithm of $G(u,v)$ is a symmetric function of $\tanh(\pi u)$ and $\tanh(\pi u)$, with an explicit factor of $(\tanh(\pi u)-\tanh(\pi u))^2$ at every loop order.
As such, it is not only satisfying the aforementioned constraints but also $G(u+2i,v)=G(u,v)$.\footnote{This property looks like a perturbative counterpart of crossing, but we are not aware of a relation between these analytic continuations.} We assume that this property continues to hold at any order in perturbation theory.

In fact, we can make a finite-coupling conjecture for $G$ that reproduces \eqref{eq: perturbative G} by introducing building blocks quite similar to those occurring in the S-matrices, pentagon transitions and measures.

Let us begin by reviewing the key building blocks occurring in the construction of the aforementioned quantities, namely the Beisert-Eden-Staudacher (BES) kernel, the source terms $\kappa_{n}$ and $\tilde{\kappa}_{n}$ as well as the functions $f_i$ for $i=1,2,3,4$; see also appendix \ref{S-matrixapp}.
The BES kernel can be written as \cite{Beisert:2006ez}
\begin{equation}
\label{eq: BES kernel}
 \mathbb{K}_{ij}=2j(-1)^{ij+j}\int\limits_0^\infty\frac{dt}{t}\frac{J_i(2gt)J_j(2gt)}{e^t-1}\,,
\end{equation}
where $J_i$ denotes a Bessel function.
The functions $f_{1,2,3,4}$, introduced in \cite{Basso:2013pxa}, are constructed by contracting the source terms $\kappa$ and $\tilde\kappa$,
\begin{equation}
\label{eq: source terms}
 \begin{aligned}
  \kappa_n^v=-\int\limits_0^\infty\frac{dt}{t}J_n(2gt)\frac{\cos(vt)e^{t/2}-J_0(2gt)}{e^t-1}\,,\qquad
  \tilde{\kappa}_n^v=-\int\limits_0^\infty\frac{dt}{t}J_n(2gt)\frac{\sin(vt)e^{t/2}}{e^t-1}\,,
 \end{aligned}
\end{equation}
with the BES kernel (\ref{eq: BES kernel}) in the following way:
\begin{equation}
\label{eq: f1 f2 f3 f4}
\begin{aligned}
 f_1(u,v)=2\,\tilde\kappa^u\,\mathbb{Q}\,\frac{1}{1+\mathbb{K}}\,\kappa^v\,,\qquad
 f_2(u,v)=2\,\kappa^u\,\mathbb{Q}\,\frac{1}{1+\mathbb{K}}\,\tilde\kappa^v\,,\\
 f_3(u,v)=2\,\tilde\kappa^u\,\mathbb{Q}\,\frac{1}{1+\mathbb{K}}\,\tilde\kappa^v\,,\qquad
 f_4(u,v)=2\,\kappa^u\,\mathbb{Q}\,\frac{1}{1+\mathbb{K}}\,\kappa^v\,, 
\end{aligned}
\end{equation}
where $\mathbb{Q}_{ij}=(-1)^{i+1}\,\delta_{ij}\,i$.
Note that at finite coupling the source terms $\kappa$ and $\tilde\kappa$ are vectors with infinitely many components, and $\mathbb{K}$ is an infinite matrix acting on them. At any loop order at weak coupling, only finitely many of these components are non-vanishing, making this representation particularly suitable for a weak-coupling analysis.

Next, we define the even and odd parts of the source terms $\kappa_{n}$ and $\tilde{\kappa}_{n}$ \eqref{eq: source terms}:
\beq\begin{aligned}
  \kappa_{n}^v&=\kappa_{+,n}^v+\kappa_{-,n}^v
  \,,
  &&
  \tilde{\kappa}_{n}^v&=\tilde{\kappa}_{+,n}^v+\tilde{\kappa}_{-,n}^v
  \,,
\end{aligned}
\eeq
where the odd (even) part vanishes for even (odd) $n$.
Since the  weak-coupling expansion of the source terms $\kappa_{+,n}^v$ and $\tilde{\kappa}_{-,n}^v$ is periodic under $v\to v+2i$ and can be expressed in terms of $\tanh(v\pi)$, $\kappa_{+,n}^v$ and $\tilde{\kappa}_{-,n}^v$ are natural building blocks for $G$, while $\kappa_{-,n}^v$ and $\tilde{\kappa}_{+,n}^v$ are not.
Moreover, the BES kernel \eqref{eq: BES kernel}
can be generalized to a tilted BES kernel by splitting it into four blocks as \cite{Basso:2020xts}
\begin{equation}\label{tiltedkernel}
\mathbb{K}=\mathbb{K}(\pi/4)\,,\qquad  \mathbb{K}(\alpha)=2\cos(\alpha)\begin{pmatrix}
             \cos(\alpha)\mathbb{K}_{\circ\circ} &\sin(\alpha)\mathbb{K}_{\circ\bullet}\\
            \sin(\alpha) \mathbb{K}_{\bullet\circ} &\cos(\alpha)\mathbb{K}_{\bullet\bullet}
            \end{pmatrix}\,.
\end{equation}
Here, $\mathbb{K}_{\circ\circ}$ is the overlap of odd Bessel functions with odd Bessel functions, $\mathbb{K}_{\circ\bullet}$ is the overlap of odd Bessel functions with even Bessel functions, etc.
For $\alpha=\pi/4$, one recovers the original BES kernel. For $\alpha=0$, the kernel does not mix between even and odd parts and is therefore the natural kernel to consider for our problem. It has previously occurred for the origin of the hexagon \cite{Basso:2020xts} as well as the octagon \cite{Coronado:2018ypq,Coronado:2018cxj,Kostov:2019stn,Kostov:2019auq,Belitsky:2019fan,Bargheer:2019exp}, and 
is referred to as the octagon kernel.
Using it, we now define functions $f_5$ and $f_6$ in analogy to \eqref{eq: f1 f2 f3 f4}:%
\footnote{One might also define
\begin{equation*}\label{f7f8}
 f_7(u,v)=2\,\kappa_-^u\,\mathbb{Q}\,\frac{1}{1+\mathbb{K}(0)}\,\kappa_-^v\,,\qquad
 f_8(u,v)=2\,\tilde{\kappa}_+^u\,\mathbb{Q}\,\frac{1}{1+\mathbb{K}(0)}\,\tilde{\kappa}_+^v\,.
\end{equation*}
As previously mentioned, they are not symmetric under $u\to u+2i$ and $v\to v+2i$, and hence they do not play a role for the FF transition under consideration. However, it is tempting to speculate that they could occur in other quantities. 
}
\begin{equation}\label{f5f6}
 f_5(u,v)=2\,\kappa_+^u\,\mathbb{Q}\,\frac{1}{1+\mathbb{K}(0)}\,\kappa_+^v\,,\qquad
 f_6(u,v)=2\,\tilde{\kappa}_-^u\,\mathbb{Q}\,\frac{1}{1+\mathbb{K}(0)}\,\tilde{\kappa}_-^v\,.
\end{equation}
These functions are symmetric in $u$ and $v$ and vanish for $g=0$.

We can use the functions $f_5$ and $f_6$ to build $G(u,v)$ that matches \eqref{eq: perturbative G} as well as the weak-coupling data up to eight-loop order, while satisfying all the constraints imposed on it:
\begin{equation}\label{Gf5f6}
G(u,v)=\exp\left[2f_5(u,v)-f_5(u,u)-f_5(v,v)-2f_6(u,v)+f_6(u,u)+f_6(v,v)\right]\,.
\end{equation}
This solution is manifestly symmetric in $u$ and $v$ and equal to $1$ for $g=0$ and for $u=v$. To fully prove its validity, in appendix \ref{app: crossing symmetry} we show that it also has the right behaviour under crossing, namely $G(u^{2\gamma},v)=-\,G(u,v)$.

Inserting the various finite-coupling expressions into \eqref{eq: scalar F in terms of G} and taking the square root, we find the following expression for the scalar FF transition:
\begin{align}
 \Fphiphibar(u,v)=-\,&\frac{4\,\Gamma\left(i (u-v)\right) }{g^2\,(u-v-i)\,(u-v-2 i)\,\Gamma \left(\frac{1}{2}+i u\right) \Gamma \left(\frac{1}{2}-i v\right)}
 \nn\\
 \times\exp\Big[&\int\limits_0^\infty \frac{dt}{t}\frac{ J_0(2 g t)^2-e^{t/2} [J_0(2 g t)-1] \left(e^{-i t u}+e^{i t v}\right)-1}{e^t-1}
 \label{eq: scalar two-particle at finite coupling}
 \\
 &-i f_1(u,v)+i f_2(u,v)-\tfrac{1}{2} f_3(u,u)-\tfrac{1}{2} f_3(v,v)+\tfrac{1}{2} f_4(u,u)+\tfrac{1}{2} f_4(v,v)\nn\\
&+2f_5(u,v)-f_5(u,u)-f_5(v,v)-2f_6(u,v)+f_6(u,u)+f_6(v,v)\Big]\nn
 \,,
\end{align}
where the first line reproduces \eqref{eq: scalars at Born level} at Born level and the second and third line stem from the S-matrix \eqref{SmatrixFC} and the measure \eqref{measureFC}.

\subsection{Strong coupling}\la{SCsection}

The solution (\ref{Gf5f6}) matches highly non-trivial data up to eight loops. We now show that it also agrees with the minimal solution to the bootstrap axioms at strong coupling.\footnote{We thank B. Basso for valuable discussions and for sharing his note on twist operators in the $O(N)$ $\sigma$-model.}

At strong coupling, the scalars are the lightest excitations. They describe the fluctuations of the GKP string on the sphere, and their dynamics are governed by the $O(6)$ non-linear $\sigma$-model. This leads us to interpret the two-scalar FF transition at strong coupling as a two-dimensional form factor of an operator in the $O(6)$ $\sigma$-model (see \cite{Basso:2014jfa} for a similar identification of the pentagon transition):
\beq\la{Phidefinition}
\<\Phi(0)|\theta_1,\theta_2\>_{i_1i_2}\equiv \lim_{g\to\infty}\left.F_{\phi_i\phi_j}(u_1,u_2)\right|_{u_i=\frac{2}{\pi}\theta_i}\,.
\eeq
Here, $\theta_i$  
are the hyperbolic rapidities parametrizing the incoming scalars' relativistic dispersion relation. The scalar indices are mapped to the $O(6)$ polarizations of the two incoming scalar excitations.

Through (\ref{Phidefinition}), the bootstrap axioms of the FF transition are translated into defining properties of the operator $\Phi$. Firstly, the singlet axiom together with the relativistic symmetry of the $O(6)$ $\sigma$-model implies that 
\beq
\<\Phi(0)|\theta_1,\theta_2\>_{i_1i_2}={\mathbb F}(\theta_1-\theta_2)\,\delta_{i_1i_2}\,,
\eeq
where ${\mathbb F}(\theta)$ is the strong-coupling limit of $\left.\Fphiphibar(u,v)\right|_{u-v={2\over\pi}\theta}$. The crossing axiom (\ref{eq: crossing}) now takes the form
\beq\la{SCcrossing}
{\mathbb F}(\theta+i\pi)={\mathbb F}(-\theta)\,.
\eeq
This implies that $\Phi(0)$ is a twist field which generates a conical defect angle of $\pi$ around the origin.\footnote{This picture is also supported by the minimal area solution dual to the $n$-gluon form factor \cite{Maldacena:2010kp,Gao:2013dza}. In the collinear limit, the induced metric on the worldsheet approaches $ds^2\simeq\[P(z)\bar P(z)\]^{1\over4}dz\,d\bar z$, with $P(z)={1\over z^2}\prod_{i=1}^{n-2}(z-z_i)$. Due to the $1/z^2$ behavior near the origin, there is a conical deficit angle of $\pi$ that corresponds to the FF transition. Similarly, around the points $z_i$ there is a conical excess angle of $\pi/2$ that corresponds to the $n-2$ pentagon transitions in (\ref{FFOPE}).} Secondly, the form factor ${\mathbb F}(\theta)$ has to satisfy the Watson equation
\beq\label{WatsonEq}
{\mathbb F}(\theta) = S_{\textrm{singlet}}(\theta)\times{\mathbb F }(-\theta)\,,
\eeq
where the singlet S-matrix is given by \cite{Zamolodchikov:1977nu}
\beq\la{SCS}
S_{\textrm{singlet}}(\theta)=\frac{\Gamma\left(\ft{3}{2}-i\ft{\theta}{2\pi}\right)\Gamma\left(\ft{5}{4}-i\ft{\theta}{2\pi}\right)\Gamma\left(\ft{3}{4}+i\ft{\theta}{2\pi}\right)\Gamma\left(+\,i\ft{\theta}{2\pi}\right)}{\Gamma\left(\ft{3}{2}+i\ft{\theta}{2\pi}\right)\Gamma\left(\ft{5}{4}+i\ft{\theta}{2\pi}\right)\Gamma\left(\ft{3}{4}-i\ft{\theta}{2\pi}\right)\Gamma\left(-\,i\ft{\theta}{2\pi}\right)}\,.
\eeq
It can also be obtained by taking the strong-coupling limit of $S_{\phi\bar\phi}(u,v)$ from (\ref{SmatrixFC}), with $\theta=\ft{\pi}{2}(u-v)$.

Lastly, the square-limit pole (\ref{slscalar}) takes the form
\beq\la{SCsqurelimit}
\lim_{\theta\to0}{\mathbb F}(\theta) =
{i\pi\over\mu}
\times{1\over\theta}\,,
\eeq
where $\mu=g\,{\sqrt\pi\,\Gamma(3/4)\over\Gamma(1/4)}$ is the strong-coupling limit of the scalar measure. Under crossing (\ref{SCcrossing}), this square-limit pole maps into the standard kinematical pole of two-dimensional relativistic form factors:
\beq\label{kine}
{\mathbb F}(\theta+i\pi) \sim \frac{1}{i\theta}\,.
\eeq

Following \cite{Karowski:1978vz,Cardy:2007mb} we construct a minimal solution to (\ref{SCcrossing})-(\ref{SCsqurelimit}) that is analytic in the physical strip $0<\Im m(\theta) <2\pi$, except for the square-limit pole (\ref{SCsqurelimit}) and its images under crossing (\ref{SCcrossing}). Note first that the scattering phase (\ref{SCS}) has the integral representation~\cite{Karowski:1978vz}
\beq\label{Ssinglet}
S_{\textrm{singlet}}(\theta)= -\exp\left[-i\int\limits_0^\infty{dt\over t} g(t)\sin(t\ft{\theta}{\pi})\right]\,,\qquad g(t) = 2\,\frac{e^{-t/2}+e^{-t}}{e^{t}+1}\,.
\eeq
Using this representation, the minimal solution to the twist field matrix elements takes the form \cite{Cardy:2007mb}
\beq\label{minimalF}
{\mathbb F}(\theta) = \frac{2i\pi}{g\sinh\theta}\times \exp{\left[\int\limits_0^\infty{dt\over t}\, g(t){\sin^2\(t(\ft{\theta}{2\pi}-\ft{i}{4})\)\over \sinh(\ft{t}{2})}\right]}\,.
\eeq
Here, the factor of $1/\sinh\theta$ accounts for the kinematical and square-limit poles, as well as the minus sign in the right-hand side of~(\ref{Ssinglet}), while the exponential factor produces the singlet S-matrix in (\ref{WatsonEq}) without introducing any other singularity. By performing the integral explicitly, we arrive at
\beq
\label{eq: strong coupling FFT result}
{\mathbb F}(\theta)=\frac{i \pi\,\Gamma\({1\over4}\)\Gamma\({5\over4}\)\,\text{csch}(\theta )}{2g\,\Gamma \left(\frac{3}{4} - \frac{i\theta}{2\pi}\right) \Gamma\left(1 - \frac{i\theta}{2\pi}\right) \Gamma\left(\frac{5}{4} + \frac{i\theta}{2\pi}\right) \Gamma\left(\frac{3}{2} + \frac{i\theta}{2\pi}\right)}\,.
\eeq

By inserting this expression into (\ref{eq: scalar F in terms of G}), together with the strong-coupling expressions for the S-matrix (\ref{SCS}) and the measure $\mu$, we find
\beq\la{SCG}
\lim_{g\to\infty}G(u,v)=\cosh\(\ft{\pi}{2}(u-v)\)\,.
\eeq
In appendix \ref{app: strong-coupling limit}, we confirm that this is indeed the strong-coupling limit of our finite-coupling ansatz (\ref{Gf5f6}). Hence, we are confident that this ansatz is correct and, in particular, that it does not miss any non-perturbative contributions to the CDD factor $G(u,v)$.

Using (\ref{SCG}), it is easy to see that $G(u^{2\gamma},v)=-\,G(u,v)$. It implies that the minimal solution (with $G(u,v)=1$) also acquires a minus sign under crossing 
and therefore was not a true solution to crossing.

\section{Multi-particle form factor transitions}
\label{sec: factorization}

As briefly discussed above, the multi-particle FF transitions are expected to factorize into products of two-particle FF transitions. Here, we provide a factorized ansatz that is consistent with all the axioms and test it against the perturbative data.

\subsection{Scalars}

Let us first consider transitions involving only scalars. According to the singlet axiom, the scalars must form an $SU(4)_R$ singlet. Hence, it is only possible to have an even number of them, $2n$. As will soon become apparent, a single factorized term is not consistent with both the Watson and the square-limit axioms. Instead, we are forced to consider a sum of factorized terms. Each of them captures a subset of the square limits and is invariant under a subset of Watson relations. Different terms in the sum are related to each other by Watson equations \eqref{eq: Watson} that do not leave them invariant.
Concretely, the sum takes the form 
\beq\la{dynxmatrix}
\begin{aligned}
F_{\phi_{i_1}\dots\,\phi_{i_{2n}}}(u_1,\ldots,u_{2n})=&\,\,\Fphiphibarn(u_1,\dots,u_{2n})\times\Pi_{i_1\,\dots\,i_{2n}}(u_1,\ldots,u_{2n})\\
+&
\sum\limits_{\sigma}\Fphiphibarn^\sigma(u_1,\dots,u_{2n})\times\Pi^\sigma_{i_1\,\dots\,i_{2n}}(u_1,\ldots,u_{2n})\,,
\end{aligned}
\eeq
where $\phi_{i_1}\ldots\phi_{i_{2n}}$ are the $2n$ ordered scalars. 
Here, the terms are factorized into a dynamical part, $\Fphiphibarn$, and a matrix part, $\Pi$, that encodes the $SU(4)_R$ structure and is independent of the coupling.
The sum in the second line goes over all inequivalent structures that are generated from $\Fphiphibarn$ via Watson relations, and which can hence be labelled by permutations.%
\footnote{What exact permutations leave $\Fphiphibarn$ invariant depends, of course, on $\Fphiphibarn$, which is constructed below.}
For example, consider the permutation of the first two scalars, $\sigma_{12}$. The corresponding term in (\ref{dynxmatrix}) is
\beq\la{Fpermutation}
\begin{aligned}
&\Fphiphibarn^{\sigma_{12}}(u_1,u_2,\dots,u_{2n})\times\Pi^{\sigma_{12}}_{i_1i_2\,\dots\, i_{2n}}(u_1,u_2,\ldots,u_{2n})\\
\equiv&\[\Fphiphibarn(u_2,u_1,\dots,u_{2n})\times\Pi_{j_1j_2\,\dots\, i_{2n}}(u_2,u_1,\ldots,u_{2n})\]\times S^{j_2j_1}_{i_1i_2}(u_2,u_1)\\
=&\[\Fphiphibarn(u_2,u_1,\dots,u_{2n})\,S_{\phi\bar\phi}(u_2,u_1)\]\times\[\Pi_{j_1\,j_2\,\dots\, i_{2n}}(u_2,u_1,\ldots,u_{2n}) R^{j_2j_1}_{i_1i_2}(u_2,u_1)\]\,,
\end{aligned}
\eeq
where we factorized the S-matrix into its coupling-dependent (singlet) part and the R-matrix,
\beq\la{SR}
S_{ij}^{kl}(u,v)=S_{\phi\bar{\phi}}(u,v)\,R_{ij}^{kl}(u,v)\,.
\eeq
The explicit expressions for these factors are given in appendix \ref{S-matrixapp}. Note that (\ref{Fpermutation}) is not equivalent to just flipping the rapidities and indices in the first term of (\ref{dynxmatrix}).

\subsubsection{Dynamical part}

We would like to construct an ansatz in which the dynamical parts of all the terms in the sum in (\ref{dynxmatrix}) factorize into products of two-particle scalar transitions, $F_{\phi\bar\phi}(u,v)$ in (\ref{twoscalars}). To ensure the closure of such a factorized ansatz under the Watson relations, each term in (\ref{dynxmatrix}) has to contain either $F_{\phi\bar\phi}(u_i,u_j)$ or $1/F_{\phi\bar\phi}(u_j,u_i)$ for each pair of excitations $i<j$. 
Because $\Fphiphibar(u,v)=S_{\phi\bar\phi}(u,v)\,\Fphiphibar(v,u)$ and $S_{\phi\bar\phi}(u,v)\,S_{\phi\bar\phi}(v,u)=1$, when two neighbouring excitation are swapped, these factors produce the dynamical part of the S-matrix appearing in (\ref{Fpermutation}). It means that in total, each scalar appears in $(2n-1)$ two-particle factors, one for each other scalar. 

\begin{figure}[t]
\centering
\includegraphics[width=.98\textwidth]{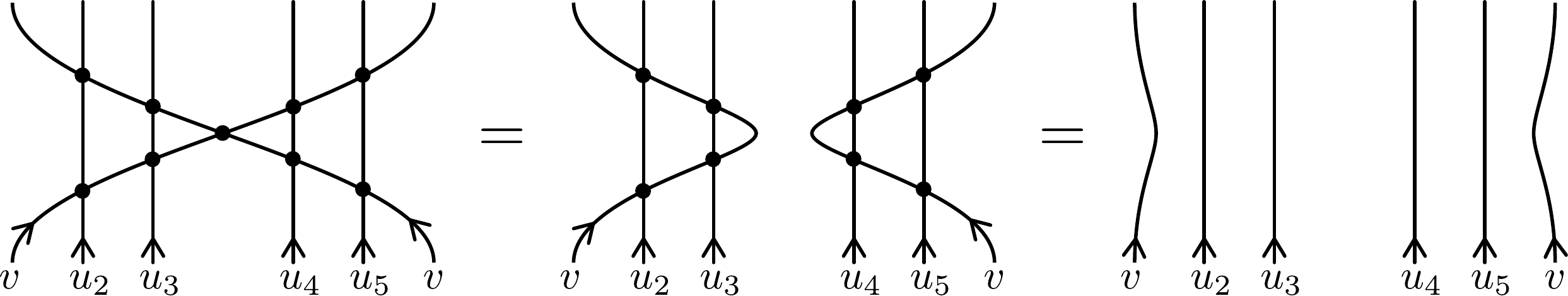}
\caption{\small The scalar S-matrix at equal rapidities is a permutation (\ref{Spermutation}). As a result, the S-matrices in the second term in the square-limit axiom (\ref{eq: squarelimit}) cancel in pairs, $S(u_i,v)S(v,u_i)=\One$, where $u_1=u_n=v$.}
\label{Smatrixcancellation}
\end{figure}

Next, we consider the square-limit axiom (\ref{eq: squarelimit}). 
It consists of two terms, one with and the other without a product of S-matrices. This might naively contradict a factorized ansatz. It turns out, however, that the product of S-matrices is trivial and the right-hand side of the square-limit axiom does factorize.
The reason for this simplification is that the S-matrix between two scalars of the same rapidity is a permutation,
\beq\la{Spermutation}
S_{ij}^{kl}(v,v)=\delta_i^l\delta_j^k\,.
\eeq
As a result, the S-matrices in the second term in (\ref{eq: squarelimit}) cancel in pairs, see figure \ref{Smatrixcancellation}. We conclude that, as for two scalars,  the two terms in the square-limit axiom combine into a principle-part prescription for the factorization pole, times the measure and the $(2n-2)$-particle FF transition.

Let us now use the Watson relation to move excitation $u_i$ to the first position, excitation $u_j$ to the last, and then take the $u_i\to u_j$ square limit. To correctly reproduce the square-limit pole, there must be a term in (\ref{dynxmatrix}) with a factor of $\Fphiphibar(u_i,u_j)$ in the numerator. For the rest of the FF transition to factor out, all other two-particle transitions involving $u_i$ and $u_j$ must cancel. Hence, among the $2n-1$ two-particle transitions involving $u_i$, $n$ factors are in the numerator and $(n-1)$ factors are in the denominator. 

One immediate consequence of this is that a single factorized term cannot be permutation invariant, as claimed above. Namely, because some of the transitions are in the numerator of $\Fphiphibarn$ and some are in the denominator, reordering the scalars using the Watson relation generically leads to a different function of the ordered scalars. This is the reason the ansatz (\ref{dynxmatrix}) involves a sum of factorized terms instead of a single one, which would have required the two-particle transitions to be all together in the numerator or denominator. This structure is more complicated than the one for the pentagon transition, for which a single factorized term is consistent with all the axioms \cite{Basso:2013vsa}.%
\footnote{The reason is that for the pentagon transition, the square-limit axiom involves excitations in two different groups, the ones on the top and the ones on the bottom of the pentagon. On the other hand, for the FF transition a single group of excitations is subject to an analogous square-limit constraint.}

\begin{figure}[t]
\centering
\includegraphics[width=15cm]{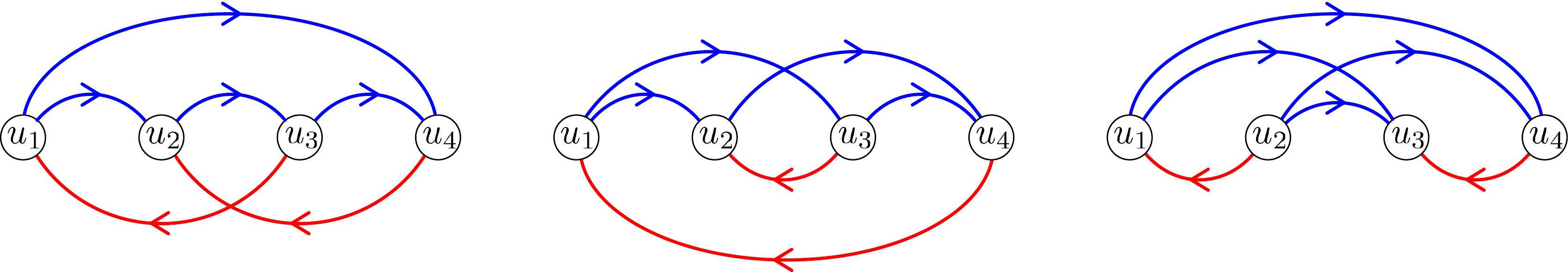}
\caption{\small The three dynamical structures entering the four-scalar FF transition, (\ref{fourscalars}). Each of them factorizes into six two-particle FF transitions. 
A blue line from $u_i$ to $u_j$ represents a factor of $F_{\phi\bar\phi}(u_i,u_j)$, while a red one represents a factor of $1/F_{\phi\bar\phi}(u_i,u_j)$.  
The first structure on the left is invariant under crossing, while the other two are interchanged.}
\label{4ptStructures}
\end{figure}

Let us illustrate this construction for the case of the four-scalar FF transition. In this case, there are three different factorized dynamical structures that are consistent with the Watson and the square-limit axioms, see figure \ref{4ptStructures}. Correspondingly, equation (\ref{dynxmatrix}) becomes
\beq\la{fourscalars}
\begin{aligned}
F_{\phi_{i_1}\phi_{i_2}\phi_{i_3}\phi_{4}}(u_1,u_2,u_3,u_4)&=\Fphiphibarfour(u_1,u_2,u_3,u_4)\times\Pi_{i_1i_2i_3i_4}(u_1,u_2,u_3,u_4)\\
&+\Fphiphibarfour^{\sigma_{34}}(u_1,u_2,u_3,u_4)\times\Pi^{\sigma_{34}}_{i_1i_2i_3i_4}(u_1,u_2,u_3,u_4)\\
&+\Fphiphibarfour^{\sigma_{23}}(u_1,u_2,u_3,u_4)\times\Pi^{\sigma_{23}}_{i_1i_2i_3i_4}(u_1,u_2,u_3,u_4)\,,
\end{aligned}
\eeq
where
\beq
\Fphiphibarfour(u_1,u_2,u_3,u_4)=\frac{\Fphiphibar(u_1,u_2)\Fphiphibar(u_2,u_3)\Fphiphibar(u_3,u_4)\Fphiphibar(u_1,u_4)}{\Fphiphibar(u_3,u_1)\Fphiphibar(u_4,u_2)}\,.
\eeq
These three structures form a closed set under Watson relations. We see that the $u_1\to u_4$ square limit is captured by the terms in the first and last lines of (\ref{fourscalars}). In this limit, the factor $\Fphiphibar(u_1,u_4)$ produces the factorization pole times the scalar measure, while the rest of the factors involving $u_1$ and $u_4$ cancel out. Similarly, the factorization pole involving $u_1$ and $u_3$ is reproduced by the second and third lines.

For the transition of $2n$ scalars we take 
the dynamical part of the first term in (\ref{dynxmatrix}) to be the ratio between the product of two-particle transitions of scalars that are separated by an odd number of sites and the product of the ones between scalars that are separated by an even number of sites
\beq\la{dynamical}
\Fphiphibarn(u_1,\dots,u_{2n})=\prod\limits_{i<j}^{2n} \frac{\Fphiphibar(u_i,u_j)^{\frac{1-(-1)^{i-j}}{2}}}{\Fphiphibar(u_j,u_i)^{\frac{1+(-1)^{i-j}}{2}}}\,.
\eeq
This term is then multiplied by the matrix part and summed over the ${1\over2}\binom{2n}{n}$  permutations that produce the inequivalent structures $\Fphiphibarn^\sigma$, see (\ref{dynxmatrix}). 

Lastly, we consider the crossing axiom. Let us focus on (\ref{dynamical}) first. The product of the two-particle FF transition in the numerator of (\ref{dynamical}) is manifestly crossing symmetric. This is because $\Fphiphibar(u_1^{2\gamma},u_j)=\Fphiphibar(u_j,u_1)$ and two scalars that are separated an odd (even) number of sites remain separated an odd (even) number of sites after crossing. Up to a simple rational factor, the denominator is also crossing invariant:
\beqa\la{crossden}
&&\Fphiphibar(u_j,u_1^{2\gamma})=S_{\phi\bar\phi}(u_j,u_1^{2\gamma})\,\Fphiphibar(u_1^{2\gamma},u_j)=S_{\phi\bar\phi}(u_j,u_1^{2\gamma})\,\Fphiphibar(u_j,u_1)\\
&&\qquad=S_{\phi\bar\phi}(u_j,u_1^{2\gamma})\,S_{\phi\bar\phi}(u_j,u_1)\,\Fphiphibar(u_1,u_j)=\frac{(u_1-u_j+3i)\,(u_1-u_j+4i)}{(u_1-u_j-i)\,(u_1-u_j-2 i)}\,\Fphiphibar(u_1,u_j)\,,\nn
\eeqa
where  we used \eqref{eq: singlet in terms of identical S-matrix} and \eqref{eq: crossing of S matrix}. 
The rational coupling-independent pre-factor on the right-hand side of (\ref{crossden}) can be absorbed in the crossing transformation of the matrix part, which we turn to next. It is not to hard to see that crossing of any of the terms in (\ref{dynxmatrix}) is equivalent to a cyclic permutation and, hence, the sum is invariant.

\subsubsection{Matrix part}

The bootstrap axioms for the FF transition reduce to simplified constraints on the matrix part. We list them here and solve them explicitly for the case of four scalars.

\paragraph{Watson}
The dynamical part satisfies the Watson relation with the dynamical part of the S-matrix, see (\ref{SR}) and (\ref{eq: Watson special}). Hence, the matrix part should satisfy the same relation, but with the R-matrix instead: 
\beq\la{WatsonMat}
\Pi^{\sigma_{ii+1}\cdot\tilde\sigma}_{\ldots\, ii+1\,\ldots}(\ldots,u_i,u_{i+1},\ldots)=\Pi^{\tilde\sigma}_{\ldots \,k+1k\,\ldots}(\ldots,u_{i+1},u_i,\ldots)\times R^{kk+1}_{ii+1}(u_{i+1},u_i)\,.
\eeq
Using this relation, the matrix parts that correspond to terms with non-trivial permutations in (\ref{dynxmatrix}) can be expressed in terms of the matrix part that multiplies the unpermuted dynamical part (\ref{dynamical}).

\paragraph{Reflection} The dynamical factor (\ref{dynamical}) is invariant under reflection (\ref{eq: reflection}). Hence, so is the matrix part
\beq\la{refMat}
\Pi_{i_1\,\dots\,i_{2n}}(-u_1,\dots,-u_{2n})=\Pi_{i_{2n}\,\dots\,i_1}(u_{2n},\dots,u_1)\,.
\eeq

\paragraph{Crossing} The crossing transformation of the factorized dynamical part (\ref{dynamical}) translates into the following crossing transformation of the corresponding matrix part 
\beq\la{crossingMat}
{\Pi_{j_1\,\dots\,j_{2n}}(u_1+2i,\dots,u_{2n})\over\Pi_{j_2\,\dots j_{2n}j_1}(u_2,\dots,u_{2n},u_1)}=\prod_{i=1}^{n-1} \frac{(u_1-u_{2i+1}+3i)\,(u_1-u_{2i+1}+4i)}{(u_1-u_{2i+1}-i)\,(u_1-u_{2i+1}-2 i)}\,,
\eeq
where we used that $u^{2\gamma}=u+2i$ and (\ref{crossden}).

\paragraph{Square limit} In the square limit, the matrix part must factorize as 
\beq\la{sqlimitMat}
\Pi_{i_1\,\dots\,i_{2n}}(u,u_2,\dots,u_{2n-1},u)={1\over2}\,\delta_{i_1i_{2n}}\times\Pi_{i_2\,\dots\,i_{2n-1}}(u_2,\dots,u_{2n-1})\,,
\eeq
where the factor of 2 in the denominator stands for the two terms in (\ref{dynxmatrix}) that reduce to the same term after removing the first and last excitations.

\paragraph{Singlet} The matrix part $\Pi_{i_1\dots i_{2n}}$ projects $2n$ scalars into an $SU(4)_R$ singlet. It consists of ${(2n-1)!!}$ tensors structures, which are factorized products of $\delta_{ij}$'s. Each one of them is multiplied by a function of the rapidities $\pi_I(u_1,\dots,u_n)$, with $I=1,\dots,(2n-1)!!$. Namely, we have 
\beq
\Pi_{i_1\,\dots\,i_{2n}}(u_1,\dots,u_{2n})=\pi_1(u_1,\dots,u_n)\prod_{k=1}^n\delta_{kk+n}+\dots\,.
\eeq

\subsubsection*{Four scalars}

We now solve for the four-scalars matrix part explicitly, starting from (\ref{fourscalars}). The first line of (\ref{fourscalars}), with the three singlet tensor structures inserted, reads
\begin{equation}\la{fourscalars2}
\begin{aligned}
\Fphiphibarfour(u_1,u_2,u_3,u_4)\,\Pi_{i_1i_2i_3 i_4}(u_1,u_2,u_3,u_4)&=\frac{\Fphiphibar(u_1,u_2)\Fphiphibar(u_2,u_3)\Fphiphibar(u_3,u_4)\Fphiphibar(u_1,u_4)}{\Fphiphibar(u_3,u_1)\Fphiphibar(u_4,u_2)}\\&\times\(\pi_1\,\delta_{i_1i_4}\delta_{i_2i_3}+\pi_2\,\delta_{i_1i_3}\delta_{i_2i_4}+\pi_3\,\delta_{i_1i_2}\delta_{i_3i_4}\)\,.
\end{aligned}
\end{equation}

In general, the Watson constraint (\ref{WatsonMat}) mixes the $\Pi$'s of the three different dynamical structures in (\ref{fourscalars}). Among all possible permutations of the four scalars, which are generated by multiple uses of the Watson relation, it is sufficient to consider the ones that take $\Pi$ in (\ref{fourscalars}) back to itself. The rest can then be used to generate the two other matrix parts in (\ref{fourscalars}) from $\Pi$. The permutations that preserve $\Pi$ are cyclic permutations, reflection, as well as any combination of these.

The crossing constraint (\ref{crossingMat}) now takes the form
\beq
\begin{aligned}
&\pi_1(u_1+2i,u_2,u_3,u_4)=\frac{(u_1-u_3-i)\,(u_1-u_3-2 i)}{(u_1-u_3+3 i)\,(u_1-u_3+4 i)}\,\pi_3(u_2,u_3,u_4,u_1)\,,\\
&\pi_3(u_1+2i,u_2,u_3,u_4)=\frac{(u_1-u_3-i)\,(u_1-u_3-2 i)}{(u_1-u_3+3 i)\,(u_1-u_3+4 i)}\,\pi_1(u_2,u_3,u_4,u_1)\,,\\
&\pi_2(u_1+2i,u_2,u_3,u_4)=\frac{(u_1-u_3-i)\,(u_1-u_3-2 i)}{(u_1-u_3+3 i)\,(u_1-u_3+4 i)}\,\pi_2(u_2,u_3,u_4,u_1)\,.
\end{aligned}
\eeq
The square-limit condition (\ref{sqlimitMat}) reads
\beq
\begin{aligned}
\left.\pi_1\right|_{u_1=u_4}=\frac{1}{2}\,, \qquad \left.\pi_2\right|_{u_1=u_4}=\left.\pi_3\right|_{u_1=u_4}=0\,,
\end{aligned}
\eeq
and the reflection symmetry of $\Pi$ (\ref{refMat}) simply becomes a symmetry of the $\pi$'s  
\begin{equation}
 \pi_i(-u_4,-u_3,-u_2,-u_1)=\pi_i(u_1,u_2,u_3,u_4)\,.
\end{equation}

These constraints are easily seen to be satisfied by the following solution:
\begin{equation}\la{4scalarM}
\begin{aligned}
\pi_2(u_1,u_2,u_3,u_4)&=
\frac{(u_1-u_4)\,(u_1-u_2-2 i)\,(u_2-u_3-2 i)\,(u_3-u_4-2 i) }{4\,(u_1-u_3+i)\,(u_1-u_3+2 i)\,(u_2-u_4+i)\,(u_2-u_4+2 i)}\,,\\ 
\pi_1(u_1,u_2,u_3,u_4)&=-\,\frac{\pi_2(u_2,u_1,u_4,u_3)-\sum\limits_{i=1}^2r^{(i)}(u_2,u_1)\,r^{(i)}(u_4,u_3)\,\pi_2(u_1,u_2,u_3,u_4)}{i\left(u_1-u_2+u_3-u_4\right)r^{(2)}(u_2,u_1)\,r^{(2)}(u_4,u_3)}\,,\\
\pi_3(u_1,u_2,u_3,u_4)&=\frac{(u_1-u_3-3i)\,(u_1-u_3-4i)}{(u_1-u_3+i)\,(u_1-u_3+2i)}\,\pi_1(u_2,u_3,u_4,u_1-2i)\,,
\end{aligned}
\end{equation}
where the rational functions $r^{(i=1,2,3)}(u,v)$ represent the coefficients of the three tensor structures of the R-matrix and are given in (\ref{rfactors}). 

In subsection \ref{subsec: matching data}, we will verify that this solution is in agreement with the perturbative data. We leave the generalization of this solution to matrix parts with more than four scalars to future work.

\subsection{Gluons and Fermions}\la{ferandglue}

For gluons and fermions, a factorized ansatz means that they only appear in singlet pairs of two conjugate excitations with the same momenta, (\ref{FpsiFFT}). Such a gluon-gluon 
or fermion anti-fermion pair behaves as an effective single-particle excitation that is an $SU(4)_R\times U(1)_\phi$ singlet. Hence, constructing a factorized ansatz involving gluons and fermions is trivial. We simply divide them into pairs and take products of two-particle transitions for all the pairs. To cover all the different multi-square limits, we have to sum over all possible divisions. We consider the ordering of the excitations in which the square limits of the gluons and fermions can be taken directly. Other orderings can be obtained using the Watson equation.

For example, the four-particle FF transitions involving only gluons and fermions are
\begin{align}\label{eq: 4pt FFbar and psipsibar}
F_{F\psi_A\bar{\psi}^B\bar{F}}(u,u',v',v) &= \delta_A^B\,F_{F\bar{F}}(u,v)\,F_{\psi\bar{\psi}}(u',v')\,,\\
F_{FF\bar{F}\bar{F}}(u,u',v',v) &= F_{F\bar{F}}(u,v)\,F_{F\bar{F}}(u',v')+S_{FF}(u,u')\,F_{F\bar{F}}(u,v')\,F_{F\bar{F}}(u',v)\,,\nn\\
F_{\psi_{A}\psi_{B}\bar{\psi}^{C}\bar{\psi}^{D}}(u,u',v',v) &= \delta_{A}^{D}\delta_{B}^{C}\,F_{\psi\bar\psi}(u,v)\,F_{\psi\bar\psi}(u',v')
-S_{AB}^{CD}(u,u')\,F_{\psi\bar\psi}(u,v')\,F_{\psi\bar\psi}(u',v)\,,\nn
\end{align} 
where $F\equiv F_1$ and $\bar{F}\equiv F_{-1}$.
Note that in the mixed transitions given in the first line 
there is a unique square-limit pairing in which each type of excitation decouples in pairs. In these cases, we have chosen one of the four orderings where this limit can be taken directly, without reordering of the excitations ($F\psi_A\bar{\psi}^B\bar{F}$). The other orderings of this type (such as $\psi_AF\bar F\bar{\psi}^B$) give the same answer because the S-matrices that result from the permutations cancel each other. In the last two lines, transitions involve four excitations of the same type. In these cases, there are two different possible pairings of conjugate excitations and, correspondingly, two terms in the FF transition, which are related by the Watson equation.

Such a factorized ansatz is consistent with the square-limit axiom because, as in the scalar case, the product of S-matrices in the second term of (\ref{eq: squarelimit}) is trivial. The reason for this is, however, somewhat different and applies only to our factorized ansatz. Consider first the case in which the $u_1$ and $u_n$ excitations in (\ref{eq: squarelimit}) are two conjugate gluons or gluon bound states. Because these are abelian excitations, their scattering with any other excitation $\chi\in\{F_m,\psi_A,\bar\psi^A,\phi_i\}$ is a phase instead of a matrix. As a result of unitarity and charge conjugation symmetry, the terms in the product of S-matrices in (\ref{eq: squarelimit}) cancel in pairs,
\beq\la{Scancel}
S_{\chi F_m}(v,u)S_{F_{-m}\bar\chi}(u,v)=S_{\chi F_m}(v,u)S_{F_m\chi}(u,v)=1\,,\qquad S_{F_mF_{-m}}(u,u)=1\,,
\eeq
where $u=u_1=u_n$ in (\ref{eq: squarelimit}). The same cancellation also applies for the square limit of $\chi$ with $\bar\chi$, in which $v=u_1=u_n$. Hence, for any state the gluon and gluon-bound-state S-matrices in (\ref{eq: squarelimit}) cancel in pairs.

\begin{figure}[t]
\centering
\includegraphics[width=1\textwidth]{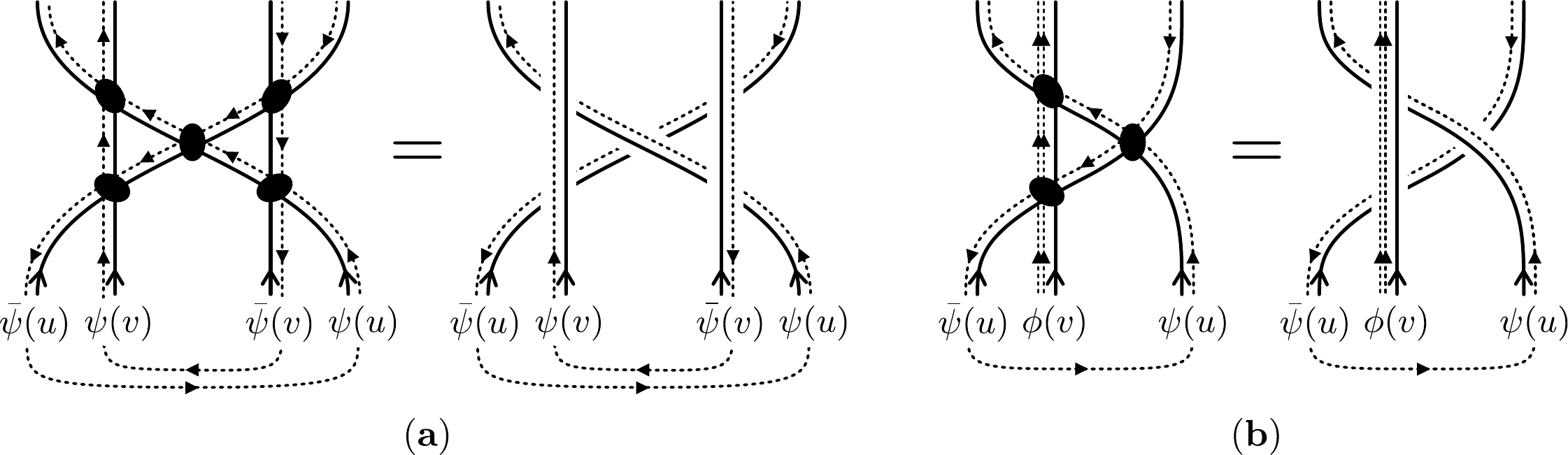}
\caption{\small The square limit axiom (\ref{eq: squarelimit}) for fermions is consistent with factorization because the product of S-matrices in the second term becomes trivial when evaluated on a factorized ansatz. To see this, we note that two conjugate fermions of equal rapidity scatter trivially with a scalar (${\bf b}$) and with another pair of conjugate fermions (${\bf a}$). Here, the dashed directed lines represent the contraction of the $SU(4)_R$ index, the solid arrow represents the flow of $U(1)_\phi$ charge in $1\over2$ units and the black dots represents the R-matrices in (\ref{Rpsipsi}) and (\ref{SandR}).}
\label{psi4}
\end{figure}
Now consider the case in which the $u_1$ and $u_n$ excitations in (\ref{eq: squarelimit}) are two conjugate fermions. Because the square-limit axiom is consistent with the Watson relation, it is sufficient to consider any particular ordering of the excitations. 
To show that the product of S-matrices in the second term in (\ref{eq: squarelimit}) is trivial for any state, it is therefore enough to show that we can freely pass the $u_1$ or $u_n$ fermionic excitation through a single scalar or through another pair of conjugate fermions with the same rapidity. 
The S-matrices factorize into a dynamical and the R-matrix parts, see (\ref{Spsipsi}), (\ref{SandR}). The dynamical parts are abelian factors. They cancel due to unitarity and charge conjugation symmetry, similarly to (\ref{Scancel}). We can therefore focus on the matrix part, which is captured by the R-matrices. Using their explicit form given in (\ref{Rpsipsi}) and (\ref{SandR}), 
we have checked that indeed, the $u_1$ and $u_n$ excitations pass freely through a scalar or another pair of fermions, 
see figure \ref{psi4}.

Note also that unlike in the scalar case, in which each dynamical part in (\ref{dynxmatrix}) factorized in terms of the two-particle transitions only, in (\ref{eq: 4pt FFbar and psipsibar}) the S-matrix explicitly appears in the last two lines. 
The two terms in $F_{FF\bar F\bar F}$ and $F_{\psi\psi\bar\psi\bar\psi}$ are supported on different kinematical limits, which describe two pairs of decoupled excitations that effectively propagate on the square. Hence, no additional interaction between the pairs is present.

\subsubsection*{Fusion}
Lastly, we note that the multi-gluon FF transitions fuse properly into the gluon-bound-state FF transitions. To see this, the FF transitions have to be contracted with the pentagon creation amplitudes and the measures, in the way they enter the FFOPE (\ref{FFOPE}), see \cite{Sever:2020jjx,Sever:2021nsq} for details. For example, we can start with the four-gluon contribution to the OPE for the three-point form factor (\ref{FFOPE}):\footnote{Compared to \cite{Sever:2020jjx,Sever:2021nsq}, here we have used the reflection symmetry to replace $F_{\bar F\bar FFF}(-u_4,-u_3,-u_2,-u_1)$ with $F_{FF\bar F\bar F}(u_1,u_2,u_3,u_4)$.}
\begin{align}
&\cW_{3,F^4}=\left.\<\cF| e^{iP\sigma-H\tau}|0\>\right|_{4-\rm 
gluons}\\
&={1\over4}\int P_{FF\bar F\bar F}(0|u_1,u_2,u_3,u_4)F_{FF\bar F\bar F}(u_1,u_2,u_3,u_4)\prod_{i=1}^4\mu_F(u_i)e^{i p_F(u_i)\sigma-E_F(u_i)\tau}{du_i\over2\pi}\nn\\
&=\frac{1}{2}\int 
{1\over P_{F\bar F}(u_1|u_1)P_{F\bar F}(u_2|u_2) P_{F\bar F}(u_1|u_2)P_{F\bar F}(u_2|u_1)}\nn\\
&\qquad\qquad\qquad\qquad\qquad\qquad\qquad\quad\times\[ {\mu_F(u_1)\mu_F(u_2)\over P_{FF}(u_1|u_2)P_{FF}(u_2|u_1)}\]\prod_{i=1}^2e^{2ip_F(u_i)\sigma-2E_F(u_i)\tau}{du_i\over2\pi}\,.\nn
\end{align}
As detailed in \cite{Basso:2014koa,Basso:2014nra}, gluons are fused into bound states by evaluating the integrand at $u_1=u^+$, $u_2=v^-$ with $u^\pm=u\pm{i\over2}$, and then taking the $v\to u$ limit. In terms of the original four rapidities in the second line, this corresponds to the limit $(u_1,u_2,u_3,u_4)\to (u^+,u^-,u^-,u^+)$. The term in the brackets in the last line is the two-gluon contribution to the hexagon. It has a pole at $v=u$, whose residue is the bound-state measure $\mu_{F_2}(u)$, \cite{Basso:2014koa,Basso:2014nra}. At the same time, the first factor becomes the creation amplitude for two conjugate bound states. Hence, in this limit we get 
\begin{align}
&i\,\underset{v=u}{\operatorname{{\rm residue}}} \,{\mu_F(u^+)\mu_F(v^-)\over P_{FF}(u^+|v^-)P_{FF}(v^-|u^+)}{1\over P_{F\bar{F}}(u^+|u^+)P_{F\bar{F}}(v^-|v^-) P_{F\bar{F}}(u^+|v^-) P_{F\bar{F}}(v^-|u^+)}{du\over2\pi}\nn\\
=\,&P_{F_2 F_{-2}}(0|u,u)\,\mu_{F_2}(u)\,{du\over2\pi}=P_{F_2F_{-2}}(0|u,w)F_{F_2F_{-2}}(u,w)\,\mu_{F_2}(u)\,\mu_{F_2}(w)\,{du\,dw\over(2\pi)^2}\,,
\end{align}
which is indeed the way the two bound states contribute to the FFOPE.

\subsection{Mixed states}

Finally, let us consider states containing all possible combinations of excitations. We find that they likewise factorize, following essentially the same rules as discussed above. 

For example, for two scalars and two gluons or fermions we have 
\begin{align}
F_{F\phi_i\phi_j\bar{F}}(u,u',v',v)&=F_{\phi\bar\phi}(u',v')F_{F\bar{F}}(u,v) \,\delta_{ij}\,,\\
F_{\psi_A\phi_i\phi_j\bar{\psi}^B}(u,u',v',v)&=F_{\phi\bar\phi}(u',v')F_{\psi \bar{\psi}}(u,v)\,\delta_{ij}\delta_{A}^B\,.\nn
\end{align}
Similarly, for FF transitions involving  
$2m$ scalars, the scalar FF transitions themselves form the building blocks of the factorization. For example,
\begin{equation}
 F_{F\phi_{i_1}\ldots\phi_{i_{2m}}\bar{F}}(v_1,u_1,\dots,u_{2m},v_2)=F_{\phi_{i_1}\ldots\phi_{i_{2m}}}(u_1,\dots,u_{2m})\times F_{F\bar{F}}(v_1,v_2)\,.
\end{equation}

To summarize, we see that for any paring of the gluons, gluon bound states and fermions, they propagate as if they where on the square, with no reference to the form factor. In particular, that means that the matrix structure of a FF transition involving scalars and any number of gluons and fermions is the same as the matrix structure of the FF transition involving only the scalars. This is a great simplification compared to the case of the pentagon transitions, which include effective transitions between different type of excitations and
in which fermions also contribute to the matrix part.

\subsection{Matching with data}
\label{subsec: matching data}

As discussed in section \ref{subsec: scalar transitions weak}, perturbative data is by now available up to eight-loop order \cite{Dixon:2020bbt,PerturbativeBootstrap2}. We find that the OPE results which follow from our conjectured multi-particle FF transitions involving four particles are in perfect agreement with the perturbative data at order $e^{-4\tau}$ in the  collinear limit. 

In \cite{Sever:2021nsq}, we already reported on a full match at order $e^{-4\tau}$ up to three loops, where only two-particle states and four-particle states with so-called small fermions contribute. We now extend this to contributions from generic four-particle states.

An important complication over the two-particle case is that both the multi-particle pentagon creation amplitudes and the multi-scalar FF transitions include non-trivial matrix parts. For example, to evaluate the four-scalar contribution to the three-gluon form factor, the two matrix structures are contracted together as
\beq\la{W4scalar}
\cW_{3,\phi^4}=\frac{1}{4!}\int\! P_{\phi^i\phi^j\phi^k\phi^l}(0|u_1,u_2,u_3,u_4)F_{\phi^i\phi^j\phi^k\phi^l}(u_1,u_2,u_3,u_4)\prod_{i=1}^4\mu_\phi(u_i)e^{i p_\phi(u_i)\sigma-E(u_i)\tau}du_i.
\eeq
To date, these matrix parts have not been worked out in full generality, not even for the pentagon transitions (see \cite{Belitsky:2016vyq} for some of them). Fortunately, all matrix parts for four particles are known; we summarize them in appendix \ref{app: pentagon matrix parts and contractions}.
For the four-scalar example (\ref{W4scalar}), $F_{\phi^i\phi^j\phi^k\phi^l}$ is the four-scalar FF transition in (\ref{fourscalars}), (\ref{4scalarM}), while the matrix part of the pentagon transition is given in \eqref{eq: four scalar pentagon matrix part}. 
Using 
the technique detailed in \cite{Sever:2021nsq}, the integral above evaluates to
\begin{align}
\cW_{3,\phi^4}=\,\,&g^8e^{-4\tau}\biggl[18e^{4\sigma}+\left(40 \sigma^3+24 \sigma^2+10 \pi ^2 \sigma+36 \sigma+2 \pi ^2-\frac{847}{36}\right)e^{6\sigma}
+\cO(e^{8\sigma})\biggr]
 \nn\\
 +\,\,&g^{10}e^{-4\tau}\biggl[
 \biggl\{\left(6 \pi ^2-\frac{1323}{4}\right)e^{4\sigma}+\biggl(60 \zeta_3 \sigma^2-72 \zeta_3 \sigma-96 \sigma^5-256 \sigma^4-\frac{320}{3} \pi ^2 \sigma^3\nn\\
 &\hphantom{+g^{10}e^{-4\tau}\biggl[
 \biggl\{}-362 \sigma^3-116 \pi ^2 \sigma^2-\frac{765 \sigma^2}{2}-15 \pi ^4 \sigma-\frac{185}{2} \pi ^2 \sigma-\frac{9463 \sigma}{18}+5 \pi ^2 \zeta_3\nn\\
 &\hphantom{+g^{10}e^{-4\tau}\biggl[
 \biggl\{}-30 \zeta_5-\frac{17 \pi ^4}{5}+\frac{13300}{27}-\frac{6851 \pi ^2}{216}\biggr)e^{6\sigma} 
 +\cO(e^{8\sigma})
 \biggr\}\\
 &\hphantom{+g^{10}e^{-4\tau}\biggl[
 \biggl\{}-\tau\biggl\{144 \sigma e^{4\sigma}+\biggl(320 \sigma^4+352 \sigma^3+120 \pi ^2 \sigma^2+264 \sigma^2+72 \pi ^2 \sigma\nn\\&\hphantom{+g^{10}e^{-4\tau}\biggl[
 \biggl\{-\tau\,\biggl\{}+\frac{34 \sigma}{9}+96 \zeta_3+\frac{8 \pi ^4}{3}+10 \pi ^2-\frac{1006}{9}\biggr)e^{6\sigma}+\cO(e^{8\sigma})\biggr\}\biggr]\nn\\
+\,\,&\cO(g^{12})\,.\nn
\end{align}
Adding all contributions, we performed a full check up to five-loop order, finding a perfect match with the perturbative data given in the ancillary files of \cite{Dixon:2020bbt}. Moreover, we have done selective checks up to eight-loop order, again finding perfect agreement with the perturbative data given in the ancillary files of \cite{PerturbativeBootstrap2}.

It would be interesting to determine the matrix parts of the FF transitions and the pentagon transitions for states with more particles and to use them for confronting our conjectures with data also at higher orders in $e^{-2\tau}$.

\section{Conclusions}
\label{sec: conclusion}

In this paper, we have bootstrapped the FF transitions of the chiral part of the stress-tensor supermultiplet in the planar maximally supersymmetric Yang-Mills theory at finite value of the 't Hooft coupling.

The form factor transitions of a GKP state vanish 
unless the excitations are pairwise conjugate.
 For transitions involving fermions, gluons, and gluon bound states, we found that the conjugate pairs decouple and move together as if they were propagating on the square. On the other hand, the transitions involving scalars are non-trivial. They are given by a certain sum of factorized ratios of two-scalar transitions, which we have determined up to a group-theoretical coupling-independent matrix part. We found that the finite-coupling behavior of the two-scalar transition is governed by the so-called octagon kernel -- a quantity that has previously occurred for other observables. The determination of the two-scalar FF transition has benefited from a very fruitful interplay with the perturbative form factor bootstrap \cite{Dixon:2020bbt,PerturbativeBootstrap2}, and our predictions at order $e^{-2\tau}$ and $e^{-4\tau}$ perfectly match with the eight-loop perturbative data.
 Moreover, the strong-coupling limit of the two-scalar FF transition agrees with the minimal solution to the form factor axioms for a twist operator in the $O(6)$ non-linear sigma model.

We conclude by listing some of the many future directions.
\begin{itemize}
\item Both the pentagon transitions and the FF transitions factorize into a known dynamical part that encodes all dependence on the coupling and a matrix part that encodes the group-theoretic dependence on the fermions' and scalars' $SU(4)_R$ indices. The matrix part has not been determined in full generality, neither for form factors nor for pentagons. Fixing the matrix parts for the  FF transition seems simpler than for the pentagon transitions, because the former only depend on the scalars, not on the fermions. 
Moreover, instead of fixing these matrix parts individually, it may be simpler to find a closed-form integral representation for the full contraction of the matrix parts that contribute to a particular $n$-point form factor or $n$-point amplitude, as was done for the case of the six-point amplitude in \cite{Basso:2015uxa}.

The missing matrix parts would allow us to perform more tests of the FFOPE and produce more high-loop predictions.

\item It would be interesting to see how the minimal surface area of \cite{Maldacena:2010kp,Gao:2013dza} emerges from the FFOPE at large coupling. In particular, it would be important to compute the scalar correction to it, which is expected to be of the same order as the minimal area \cite{Basso:2014jfa}. 

\item The octagon kernel has previously occurred in the description of four-point functions  \cite{Coronado:2018ypq,Coronado:2018cxj,Kostov:2019stn,Kostov:2019auq,Belitsky:2019fan,Bargheer:2019exp} 
and the origin of the six-point amplitude \cite{Basso:2020xts}; see also \cite{Arkani-Hamed:2021iya}. 
Very recently, the three-point form factor was related to a particular subspace of the six-point amplitude in an intricate way \cite{Dixon:2021tdw}.
It would be interesting to see whether this relation can be used to connect the occurrences of the octagon kernel in both cases. Moreover, it would be interesting to understand a potential connection with the octagon.

\item While we used the three-point MHV form factor to guide and check our derivation of the FF transitions, they equally determine the $n$-point MHV form factor. Adding more than three gluons in MHV configuration is achieved by simply gluing more OPE channels, connected by the pentagon transitions, see (\ref{FFOPE}). This should provide valuable data for a perturbative bootstrap at $n>3$.
\item 
It would be interesting to consider the N$^k$MHV configurations with $k\ge1$, which would require the known charged pentagons \cite{Basso:2014hfa,Basso:2015rta}.
\item The chiral part of the stress-tensor supermultiplet is in a sense the simplest operator multiplet, having BPS protected dimension and trivial higher integrability charges. The form factors of local operators with non-trivial charges can equally be incorporated into the FFOPE. One has to twist the crossing axiom (\ref{eq: crossing}) by (T-dual of) the corresponding charges, see \cite{Cavaglia:2020hdb}. It would be fascinating to bootstrap the corresponding FF transitions and try to express them in terms of the operators' Q-functions. Corresponding perturbative data is available in \cite{Engelund:2012re,Brandhuber:2014ica,Wilhelm:2014qua, Nandan:2014oga, Loebbert:2015ova, Brandhuber:2016fni, Loebbert:2016xkw, Caron-Huot:2016cwu,Banerjee:2016kri,Ahmed:2016vgl,Brandhuber:2018xzk,Lin:2020dyj,Guo:2021bym,Lin:2021qol}.
\item 
By inserting a complete basis of states, form factors can be used to decompose other observables in terms of sums of product of form factors. In particular, it would be interesting to decompose the double-trace corrections to scattering amplitudes. These are dual to correlation functions between two periodic Wilson loops \cite{Ben-Israel:2018ckc}, such as the ones we studied in the present paper. 
The expansion of this correlator around the limit where the two Wilson loops are far apart is controlled by the FF transition of the low-lying 
operators, the leading contribution being given by the FF transitions of the chiral part of the stress-tensor supermultiplet, which we studied in this paper. 
\end{itemize}

\paragraph{Acknowledgments:} 
We are very grateful to B.~Basso for many valuable discussions and comments on the manuscript. MW thanks L.~Dixon, {\"O}.~G{\"u}rdo{\u{g}}an and A.~McLeod for collaboration on \cite{Dixon:2020bbt,PerturbativeBootstrap2}. AT and MW are grateful to CERN for hospitality. AS is grateful to NBI for hospitality. AS was supported by the Israel Science Foundation (grant number 1197/20). AT received funding from the European Research Council (ERC) under the European Unions Horizon 2020 research and innovation programme, Novel structures in scattering amplitudes (grant agreement No. 725110).
MW was supported in part by the ERC starting grant 757978 and the research grants 00015369 and 00025445 from Villum Fonden.

\appendix

\section{Scalar S-matrix and measure}\la{S-matrixapp}

In this appendix, we review the scalar S-matrix \cite{Basso:2013pxa} as well as the scalar measure, in a form that is convenient for our needs.

The S-matrix between two scalar GKP excitations takes the form
\begin{equation}
S_{ij}^{kl}(u, v) = S_{\phi\bar{\phi}}(u,v)\,R_{ij}^{kl}(u,v)\, ,
\end{equation}
where
\begin{equation}
R_{ij}^{kl}(u,v) = r^{(1)}(u,v)\,\delta_{i}^{k}\delta_{j}^{l} +r^{(2)}(u,v)\,\delta_{i}^{l}\delta^{k}_{j} + r^{(3)}(u,v)\,\delta_{ij}\delta^{kl}\,,
\end{equation}
with
\begin{align}\la{rfactors}
&r^{(1)}(u,v)=\frac{(u-v)(u-v-2i)}{(u-v+i)(u-v+2i)}\,,\nn\\
&r^{(2)}(u,v)=-\,\frac{i(u-v-2i)}{(u-v+i)(u-v+2i)}\,,\\
&r^{(3)}(u,v)=\frac{i(u-v)}{(u-v+i)(u-v+2i)}\,.\nn
\end{align}

The scattering phase in different channels are related by a simple phase factor. In particular, the phase in the singlet and symmetric channels are related as
\begin{equation}
\label{eq: singlet in terms of identical S-matrix}
S_{\phi\bar\phi}(u,v)\equiv{1\over6}S_{ij}^{kl}(u,v)\delta^{ij}\delta_{kl}=\frac{(u-v+2i)\,(u-v+i)}{(u-v-2i)\,(u-v-i)}\,S_{\phi\phi}(u,v)\,.
\end{equation}
The S-matrix in the symmetric channel can be expressed as
\begin{equation}\label{SmatrixFC}
S_{\phi\phi}(u,v)=\frac{\Gamma\left(\tfrac{1}{2}-iu\right)\Gamma\left(\tfrac{1}{2}+iv\right)\Gamma\left(iu-iv\right)}{\Gamma\left(\tfrac{1}{2}+iu\right)\Gamma\left(\tfrac{1}{2}-iv\right)\Gamma\left(iv-iu\right)}\,\mathcal{G}(u,v)\,,
\end{equation}
where the function $\mathcal{G}(u,v)$ captures the entire dependence on the coupling constant. It is given by 
\begin{equation}
 \mathcal{G}(u,v)=\exp\biggl[
 2i \int_0^\infty\frac{dt}{t}(J_0(2gt)-1)\frac{e^{t/2}(\sin(ut)-\sin(vt))}{e^{t}-1}-2if_1(u,v)+2if_2(u,v)
 \biggr]\,,
\end{equation}
where $f_1$ and $f_2$ were defined in \eqref{eq: f1 f2 f3 f4}.

The scalar measure 
is given by \cite{Basso:2013aha}
\begin{equation}\label{measureFC}
\begin{aligned}
\mu(u)=\frac{\pi g^2}{\cosh(\pi u)}\exp\biggl[\int_0^\infty\frac{dt}{t}(J_0(2gt){-}1)\frac{2e^{t/2}\cos(ut)-J_0(2gt)-1}{e^t-1}+f_3(u,u)-f_4(u,u)\biggr],
\end{aligned}
\end{equation}
where $f_3$ and $f_4$ were defined in \eqref{eq: f1 f2 f3 f4}.

\section{Integral representations for the finite-coupling solution}
\label{app: integral representations}

In this appendix, we derive an integral representation for the functions $f_5$ and $f_6$ that enter our solution for the scalar two-particle FF transitions at  finite coupling.
The integral representations are more tractable for the analysis of crossing symmetry and strong coupling than the matrix notation, which is more advantageous at weak coupling.

Let us define, using matrix notation,
\begin{equation}
\gamma_{o,+}^v=\frac{1}{1+\mathbb{K}(0)}\kappa_+^v\,,\qquad \tilde{\gamma}_{o,-}^v=\frac{1}{1+\mathbb{K}(0)}\tilde{\kappa}_-^v\,.
\end{equation}
These objects satisfy
\begin{equation}
\gamma_{o,+}^v+\mathbb{K}(0)\,\gamma_{o,+}^v=\kappa_+^v\,,\qquad \tilde{\gamma}_{o,-}^v+\mathbb{K}(0)\,\tilde{\gamma}_{o,-}^v=\tilde{\kappa}_-^v\,,
\end{equation}
or, in components,
\begin{equation}\label{gammadefcomp}
\begin{aligned}
\gamma_{o,+,2n}^v+2\sum\limits_{m=1}^\infty\mathbb{K}_{2n,2m}\gamma_{o,+,2m}^v&=\kappa_{+,2n}^v\,,\\
\tilde{\gamma}_{o,-,2n-1}^v+2\sum\limits_{m=1}^\infty\mathbb{K}_{2n-1,2m-1}\tilde{\gamma}_{o,-,2m-1}^v&=\tilde{\kappa}_{-,2n-1}^v\,,
\end{aligned}
\end{equation}
where, similarly to (\ref{eq: source terms}), we defined
\begin{equation}\label{gammaintrep}
\gamma_{o,+,2n}=\int_0^\infty\frac{dt}{t}J_{2n}(2gt)\gamma_{o,+}^v(2gt)\,,\qquad \tilde{\gamma}_{o,-,2n-1}=\int_0^\infty\frac{dt}{t}J_{2n-1}(2gt)\tilde{\gamma}_{o,-}^v(2gt)\,.
\end{equation}
The odd components of $\gamma_{o,+}$ and the even components of $\tilde{\gamma}_{o,-}$ are absent as a direct consequence of the fact that the tilted BES kernel (\ref{tiltedkernel}) is diagonal for $\alpha=0$ and therefore does not mix even and odd components. The above relations can be inverted using the orthogonality of the Bessel functions,
\begin{equation}
\int_0^\infty \frac{dt}{t}J_{2n}(t)J_{2m}(t)=\frac{\delta_{n,m}}{4n}\,,\qquad \int_0^\infty \frac{dt}{t}J_{2n-1}(t)J_{2m-1}(t)=\frac{\delta_{n,m}}{2(2n-1)}\,,
\end{equation}
which results in
\begin{equation}\label{invgamma}
\gamma_{o,+}^v(2gt)=\sum_{n=1}^\infty2(2n)J_{2n}(2gt)\gamma_{o,+,2n}\,,\qquad\tilde{\gamma}_{o,-}^v(2gt)=\sum_{n=1}^\infty2(2n-1)J_{2n-1}(2gt)\tilde{\gamma}_{o,-,2n-1}\,.
\end{equation}

Using the definition of the BES kernel (\ref{eq: BES kernel}) and (\ref{invgamma}), we can rewrite (\ref{gammadefcomp}) in the following way
\begin{equation}
\label{eq: integral equation for gamma o plus}
\begin{aligned}
\gamma_{o,+,2n}^v+2\int_0^\infty \frac{dt}{t}\frac{J_{2n}(2gt)\gamma_{o,+}^v}{e^t-1}&=\kappa_{+,2n}^v\,,\\
\tilde{\gamma}_{o,-,2n-1}^v+2\int_0^\infty\frac{dt}{t}\frac{J_{2n-1}(2gt)\tilde{\gamma}_{o,-}^v}{e^t-1}&=\tilde{\kappa}_{-,2n-1}^v\,.
\end{aligned}
\end{equation}
These equations can be brought into the form 
\begin{equation}
\label{eq: alternative form gamma}
 \int_0^\infty \frac{dt}{t}(\cos(ut)-J_0(2gt))\left[\gamma_{o,+}^v(2gt)\,\frac{e^t+1}{e^t-1}+\frac{\cos(vt)e^{t/2}-J_0(2gt)}{e^t-1}\right]=0
\end{equation}
and
\begin{equation}
\label{eq: alternative form gamma tilde}
 \int_0^\infty \frac{dt}{t}\sin(ut)\left[\tilde\gamma_{o,-}^v(2gt)\,\frac{e^t+1}{e^t-1}+\frac{\sin(vt)e^{t/2}}{e^t-1}\right]=0\,.
\end{equation}
Both equations only hold for $u\leq(2g)^2$. 
These relations will be later used to find the strong-coupling behavior of $\gamma_{o,+}^v$ and $\tilde{\gamma}_{o,-}^v$, as well as to determine how the mirror transformation acts on $f_5$ and $f_6$ (\ref{f5f6}), which we can now rewrite in the following way
\begin{equation}\label{f5gammarep}
f_5(u,v)=\int_0^\infty\frac{dt}{t}\frac{\cos(ut)\,e^{t/2}-J_0(2gt)}{e^t-1}\,\gamma_{o,+}^v(2gt)\,,
\end{equation}
and similarly
\begin{equation}\label{f6gammarep}
f_6(u,v)=-\,\int_0^\infty\frac{dt}{t}\frac{\sin(ut)\,e^{t/2}}{e^t-1}\,\tilde{\gamma}_{o,-}^v(2gt)\,.
\end{equation}

\section{Crossing symmetry}
\label{app: crossing symmetry}

In this appendix, we check that the finite-coupling solution \eqref{eq: scalar two-particle at finite coupling} we found for the scalar FF transition indeed satisfies the crossing axiom \eqref{eq: crossing}.%
\footnote{A similar calculation for the S-matrix can be found in appendix B.2 of \cite{Basso:2013pxa}.}

The mirror transformation maps the rapidity $u$ of a scalar excitation to $u^{\pm\gamma} = u\pm i$. While this operation seems rather straightforward, one has to keep in mind that it involves passing through cuts connecting the points $-2g\pm i/2$ and $2g\pm i/2$, which would not be visible from within perturbation theory. To perform this delicate operation, we split the mirror transformation into three distinct steps: shifting $u$ all the way up to the cut, actually passing through the cut, and the remaining $\pm \frac{i}{2}$ shift that completes the mirror transformation. 

We first follow this prescription to compute the mirror transformation acting on $f_6$, using its integral representation (\ref{f6gammarep}) in terms of $\tilde{\gamma}_{o,-}^v$. We define $u^\pm=u\pm\frac{i}{2}\mp i0$ and using 
\begin{equation}
 \sin(u^\pm t)=\sin(ut)e^{-t/2}\pm ie^{\mp iut}\sinh(t/2)
\end{equation}
find
\begin{equation}
 f_6(u^\pm,v)=-\int_0^\infty\frac{dt}{t}\frac{\sin(ut)\tilde\gamma_{o,-}^v(2gt)}{e^t-1}\mp\frac{i}{2}\int_{0}^{\infty}\frac{dt}{t}e^{\mp iut}\tilde\gamma_{o,-}^v(2gt)\,.
\end{equation}
Next, we move through the cut using 
\begin{equation}
 \mp\frac{i}{2}\int_0^\infty\frac{dt}{t}e^{\mp iut}\tilde\gamma_{o,-}^v(2gt)=\mp\frac{i}{2}\int_0^\infty\frac{dt}{t}e^{\pm iut}\tilde\gamma_{o,-}^v(2gt)-\int_{0}^{\infty}\frac{dt}{t}\sin(ut)\tilde\gamma_{o,-}^v(2gt)\,,
\end{equation}
leading to 
\begin{align}
  f_6(u^\pm,v)&=-\int_0^\infty\frac{dt}{t}\frac{\sin(ut)\tilde\gamma_{o,-}^v(2gt)}{e^t-1}\mp\frac{i}{2}\int_0^\infty\frac{dt}{t}e^{\pm iut}\tilde\gamma_{o,-}^v(2gt)-\int_{0}^{\infty}\frac{dt}{t}\sin(ut)\tilde\gamma_{o,-}^v(2gt)\nn\\
  &= \int_0^\infty\frac{dt}{t}\frac{\sin(u^\mp t)e^{t/2}\tilde\gamma_{o,-}^v(2gt)}{e^t-1}-\int_0^\infty\frac{dt}{t}\sin(ut)\frac{e^t+1}{e^t-1}\tilde\gamma_{o,-}^v(2gt)\nn\\
  &= \int_0^\infty\frac{dt}{t}\frac{\sin(u^\mp t)e^{t/2}\tilde\gamma_{o,-}^v(2gt)}{e^t-1}+\int_0^\infty\frac{dt}{t}\sin(ut)\frac{\sin(vt)e^{t/2}}{e^t-1}\,,
\end{align}
where in the last step we used \eqref{eq: alternative form gamma tilde}. Performing the final shift by $\pm \frac{i}{2}$, we notice that the first term in the last line gives us back the original $f_6$. As a result, we find
\begin{equation}
\begin{aligned}
f_6(u^{\pm\gamma},v)&=- f_6(u,v)+\int_0^\infty\frac{dt}{t}\sin(u^\pm t)\frac{\sin(vt)e^{t/2}}{e^t-1}\,.
\end{aligned}
\end{equation}
Taking the difference between the $+$ and $-$ versions of this equation gives us
\begin{equation}
 f_6(u^{\gamma},v)-f_6(u^{-\gamma},v)=i\int_0^\infty\frac{dt}{t}\cos(ut)\sin(vt)\,.
\end{equation}

We now repeat the same derivation for $f_5$ using its integral representation (\ref{f5gammarep}) in terms of $\gamma_{o,+}^v$. We have 
\begin{equation}
 \cos(u^\pm t)=\cos(ut)e^{-t/2}+ e^{\mp iut}\sinh(t/2)\,.
\end{equation}
Using it, we find 
\begin{equation}
 f_5(u^\pm,v)=\int_0^\infty\frac{dt}{t}\frac{\cos(ut)-J_0(2gt)}{e^t-1}\gamma_{o,+}^v(2gt)+\frac{1}{2}\int_0^\infty\frac{dt}{t}e^{\mp iut}\gamma_{o,+}^v(2gt)\,.
\end{equation}
To cross through the cut, we use
\begin{equation}
 \frac{1}{2}\int_0^\infty\frac{dt}{t}e^{\mp iut}\gamma_{o,+}^v(2gt)=-\frac{1}{2}\int_0^\infty\frac{dt}{t}e^{\pm iut}\gamma_{o,+}^v(2gt)+\int_0^\infty\frac{dt}{t}\cos( ut)\gamma_{o,+}^v(2gt)\,.
\end{equation}
We thus have
\begin{equation}
\begin{aligned}
f_5(u^\pm,v)&
=\int_0^\infty\frac{dt}{t}\frac{\cos(ut)-J_0(2gt)}{e^t-1}\gamma_{o,+}^v(2gt)
\\&\phaneq -\frac{1}{2}\int_0^\infty\frac{dt}{t}e^{\pm iut}\gamma_{o,+}^v(2gt)+\int_0^\infty\frac{dt}{t}\cos( ut)\gamma_{o,+}^v(2gt)\\
&=-\int_0^\infty\frac{dt}{t}\frac{\cos(u^\mp t)e^{t/2}-J_0(2gt)}{e^t-1}\gamma_{o,+}^v(2gt)\\&\phaneq+\int_0^\infty\frac{dt}{t}\left(\cos(ut) - J_0(2gt)\right)\frac{e^t+1}{e^t-1}\gamma_{o,+}^v(2gt)\\
&=-\int_0^\infty\frac{dt}{t}\frac{\cos(u^\mp t)e^{t/2}-J_0(2gt)}{e^t-1}\gamma_{o,+}^v(2gt)\\&\phaneq-\int_0^\infty\frac{dt}{t}\left(\cos(ut) - J_0(2gt)\right)\frac{\cos(vt)e^{t/2}-J_0(2gt)}{e^t-1}\,,
\end{aligned}
\end{equation}
where in the second line we used that $\int_0^\infty\frac{dt}{t}J_0(2gt)\gamma_{o,+}^v(2gt)=0$ due to the orthogonality of the Bessel functions, and in the last line we have used \eqref{eq: alternative form gamma}.
The final shift results in
\begin{equation}\label{f5mirr}
f_5(u^{\pm\gamma},v)=-f_5(u,v)-\int_0^\infty\frac{dt}{t}\left(\cos(u^\pm t) - J_0(2gt)\right)\frac{\cos(vt)e^{t/2}-J_0(2gt)}{e^t-1}\,.
\end{equation}
This implies
\begin{equation}
f_5(u^{\gamma},v)-f_5(u^{-\gamma},v)=i\int_0^\infty\frac{dt}{t}\sin(u t)\cos(vt) - i\int_0^\infty\frac{dt}{t}e^{-t/2}J_0(2gt)\sin(ut)\,.
\end{equation}
Combining $f_5$ and $f_6$, we find
\begin{align}\label{crossingUV}
f_5(u^{\gamma},v)&-f_5(u^{-\gamma},v)-f_6(u^{\gamma},v)+f_6(u^{-\gamma},v)\\
&=i\int_0^\infty\frac{dt}{t}\sin((u-v)t)- i\int_0^\infty\frac{dt}{t}e^{-t/2}J_0(2gt)\sin(ut)\nn\\
&=\frac{i\pi}{2}\,{\rm sign}(u-v) - i\int_0^\infty\frac{dt}{t}e^{-t/2}J_0(2gt)\sin(ut)\,.\nn
\end{align}

Analytic continuation of $f_5(u,u)$ and $f_6(u,u)$ can be done in steps by first performing the mirror transformation in $u$, then in $v$ and then setting $v=u$. For $f_6$, this results in
\begin{equation}
f_6(u^{\pm\gamma},v^{\pm\gamma})-f_6(u,v)=\pm\,\frac{i}{2}\int_0^\infty\frac{dt}{t}\sin((u+v\pm i)t)\,,
\end{equation}
which leads to
\begin{equation}\label{f6doublemirrfin}
f_6(u^{\gamma},v^{\gamma})-f_6(u^{-\gamma},v^{-\gamma})=i\int_0^\infty\frac{dt}{t}\cosh(t)\sin((u+v)t)\,.
\end{equation}
Similarly,
\begin{align}\label{f5doublemirr}
f_5(u^{\pm\gamma},v^{\pm\gamma})-f_5(u,v)=&\pm\frac{i}{2}\int_0^\infty\frac{dt}{t}\sin((u+v\pm i)t) + \frac{1}{2}\log{g^2}\\
&-\int_0^\infty\frac{dt}{t}\left(\frac{1}{2}\,J_0(2gt)\left(e^{\pm i u^\pm t}+e^{\pm i v^\pm t}\right)-e^{-t}\right)\nn\,.
\end{align}
Deriving this equation involves a number of subtle steps, which we briefly outline here.\footnote{We thank B.~Basso for useful discussions of these steps.} After the mirror transformation in $u$ is performed, the second mirror transformation in $v$ acts on both terms in (\ref{f5mirr}). In the second term, simply replacing $v$ with $v\pm i$ would not lead to the correct result, due to this integral having a cut between $-2g$ and $2g$. One can see, however, that $v\to v\pm i$ gives the correct result for almost all terms in this integral, with an exception of the one term that comes from integrating $J_0(2gt)$ with $e^{\pm ivt}$. This integral by itself is divergent at $t=0$, so we can add a $v$-independent regulator $e^{-t}$ to it to cancel the divergence, while subtracting the same regulator from the rest of the terms. The analytic continuation through the cut can then be performed using the following relation:
\begin{align}
\int\limits_0^\infty \frac{dt}{t}\left(J_0(2gt)\,e^{\pm i(v\pm i0)t}-e^{-t}\right)= -\log{g^2}+\int\limits_0^\infty \frac{dt}{t}\left(-\,J_0(2gt)\,e^{\mp i(v\mp i0)t}+e^{-t}\right).
\end{align}
This then leads to (\ref{f5doublemirr}). The difference between the positive and negative transformations is given by
\begin{align}\label{f5doublemirrfin}
f_5(u^{\gamma},v^{\gamma})-f_5(u^{-\gamma},v^{-\gamma})&=i\int_0^\infty\frac{dt}{t}\cosh(t)\sin((u+v)t)\\
&\phaneq-i\int_0^\infty\frac{dt}{t}e^{-t/2}J_0(2gt)\left(\sin(ut)+\sin(vt)\right)\nn\,.
\end{align}
Combining (\ref{f6doublemirrfin}) with (\ref{f5doublemirrfin}) and setting $v$ equal to $u$, we find
\begin{equation}\label{crossingUU}
\begin{aligned}
f_5(u^{\gamma},u^{\gamma})-f_5(u^{-\gamma},u^{-\gamma})&-f_6(u^{\gamma},u^{\gamma})+f_6(u^{-\gamma},u^{-\gamma}) =-\,2i\int_0^\infty\frac{dt}{t}e^{-t/2}J_0(2gt)\sin(ut)\,.
\end{aligned}
\end{equation}

After inserting (\ref{crossingUV}) and (\ref{crossingUU}) into (\ref{Gf5f6}), we see that the remaining integrals cancel and conclude that
\begin{equation}
G(u^{2\gamma},v)^2 = \exp(2\pi i\,{\rm sign}(u-v))\,G(u,v)^2= G(u,v)^2\,,
\end{equation}
but
\begin{equation}
G(u^{2\gamma},v)=\exp(\pi i\,{\rm sign}(u-v))\,G(u,v)=-\,G(u,v)\,.
\end{equation}
This relation is exactly what we expect from the minimal solution at strong coupling, which is presented in section \ref{SCsection}.

\section{Strong-coupling limit}
\label{app: strong-coupling limit}

In this appendix, we find the strong-coupling limit of our finite-coupling solution for the scalar two-particle FF transition.

We start with the two equations \eqref
{eq: alternative form gamma} and \eqref
{eq: alternative form gamma tilde}.
They have the following two particular solutions:
\begin{equation}
 \gamma_{o,+,\text{part}}^v=-\frac{\cos(vt)e^{t/2}-J_0(2gt)}{e^t+1}\,,\qquad
 \tilde{\gamma}_{o,-,\text{part}}^v=-\frac{\sin(vt)e^{t/2}}{e^t+1}\,.
\end{equation}
Following \cite{Basso:2013pxa}, one would expect that the physical solutions differ from the particular solutions by a homogeneous solution that is exponentially suppressed at strong coupling and hence does not play a role in the strong-coupling limit.

For strong coupling, assuming the homogeneous solution to be suppressed, we can obtain the following simple integrals for the combinations of $f_5$ and $f_6$ that enter (\ref{Gf5f6}):
\begin{align}
2f_5(u,v)&-f_5(u,u)-f_5(v,v) = \frac{1}{2}\int\limits_0^\infty\frac{dt}{t}\frac{\left({\rm cos}(ut)-{\rm cos}(vt)\right)^2}{{\rm sinh}(t)}\nonumber\\
&= \frac{1}{2}\,{\rm log}\,\left[{\rm cosh}\left(\frac{\pi}{2}\,(u+v)\right){\rm cosh}\left(\frac{\pi}{2}\,(u-v)\right)\right] - \frac{1}{4}\,{\rm log}\left[{\rm cosh}(\pi u)\,{\rm cosh}(\pi v)\right],\\
2f_6(u,v)&-f_6(u,u)-f_6(v,v) = -\,\frac{1}{2}\int\limits_0^\infty\frac{dt}{t}\frac{\left({\rm sin}(ut)-{\rm sin}(vt)\right)^2}{{\rm sinh}(t)}\nonumber\\
&=\frac{1}{2}\,{\rm log}\left[\frac{{\rm cosh}\left(\frac{\pi}{2}\,(u+v)\right)}{{\rm cosh}\left(\frac{\pi}{2}\,(u-v)\right)}\right] - \frac{1}{4}\,{\rm log}\left[{\rm cosh}(\pi u)\,{\rm cosh}(\pi v)\right].
\end{align}
Inserting these expressions into (\ref{Gf5f6}) gives
\begin{align}
G(u,v) = {\rm cosh}\left(\frac{\pi}{2}\,(u-v)\right).
\end{align}
This is exactly what we expected from the minimal solution from the $O(6)$ non-linear sigma model, cf.\ section \ref{SCsection}.

\section{Pentagon matrix parts and contractions}
\label{app: pentagon matrix parts and contractions}

In this appendix, we collect the matrix part of the creation pentagons, which are required for the matching with data in subsection \ref{subsec: matching data}. Moreover, we provide some of the more non-trivial contractions between the matrix parts of the creation pentagon and the annihilation FF transition.
We are considering order $e^{-4\tau}$, where states of twist four occur. These are the combinations $F_{2}\bar{F}_{2}$, $FF\bar{F}\bar{F}$, $F\psi\bar{\psi}\bar{F}$, $F\phi\bar{\phi}\bar{F}$,  $\psi\psi\bar{\psi}\bar{\psi}$, $\psi\phi\bar{\phi}\bar{\psi}$ and $\phi\phi\bar{\phi}\bar{\phi}$.
Note that in addition, the fermions in these expressions can have either large or small momenta. 

\subsection{Matrix parts}

The matrix parts of the pentagon transitions that create the aforementioned states have been computed in \cite{Basso:2013aha,Belitsky:2016vyq,Sever:2021nsq}.

The matrix part arises for $SU_R(4)$-charged fundamental excitations and is trivial for pentagons involving gluons. The matrix part for a state involving two gluons and two non-gluons is identical to the one for the state involving only the two non-gluons.

The two-particle matrix parts have the following form:
\begin{align}
\Pi_{0|F\bar{F}}(0|v_1,v_2) & = 1,\\
\Pi_{0|\psi\bar{\psi}}(0|v_1,v_2) &= \frac{i}{v_1-v_2+2i},\nonumber\\
\Pi_{0|\phi\bar{\phi}}(0|v_1,v_2) & = -\,\frac{1}{(v_1-v_2+i)\,(v_1-v_2+2i)}.\nonumber
\end{align}

The matrix part for $\psi^{A_1}\psi^{A_2}\bar{\psi}_{B_1}\bar{\psi}_{B_2}$ is \cite{Belitsky:2016vyq}
\begin{align}
\label{eq: four scalar pentagon patrix part}
\Pi_{0|\psi\psi\bar{\psi}\bar{\psi}}(0|u_1,u_2,v_1,v_2)^{A_1A_2}_{B_1B_2} &= \delta^{A_1}_{B_1}\delta^{A_2}_{B_2}\,\pi_1 + \delta^{A_1}_{B_2}\delta^{A_2}_{B_1}\,\pi_2,
\end{align}
where
\begin{align}
\pi_1 &= -\,\frac{1}{\prod_{j,k=1}^2(u_j-v_k+2i)}\,\frac{(u_1-v_2+3i)\,(u_2-v_1+2i)}{(u_1-u_2+i)\,(v_1-v_2+i)}\,,\\
\pi_2 &= \frac{1}{\prod_{j,k=1}^2(u_j-v_k+2i)}\,\frac{(u_1-v_1+3i)\,(u_2-v_2+3i)- i\,(u_2-v_1)+2}{(u_1-u_2+i)\,(v_1-v_2+i)}\,.\nn
\end{align}

Similarly, for scalars, \cite{Basso:2013aha}
\begin{align}
\label{eq: four scalar pentagon matrix part}
\Pi_{0|\phi\phi\phi\phi}(0|u_1,u_2,u_3,u_4)_{i_1i_2i_3i_4} &= \delta_{i_1i_2}\delta_{i_3i_4}\,\theta_1 + \delta_{i_1i_3}\delta_{i_2i_4}\,\theta_2 + \delta_{i_1i_4}\delta_{i_2i_3}\,\theta_3,
\end{align}
where 
\begin{align}
\theta_1 &= \frac{(u_1-u_4+3i)\,(u_2-u_3+2i)\left[(u_1-u_3+2i)\,(u_2-u_4+3i)+i\,(u_3-u_4)-2\right]}{\prod_{1\leq j< k \leq 4}(u_j-u_k+i)(u_j-u_k+2i)},\nonumber\\
\theta_2 &= -\,\frac{(u_1-u_2+2i)\,(u_2-u_3+2i)\,(u_3-u_4+2i)\,(u_1-u_4+3i)}{\prod_{1\leq j< k \leq 4}(u_j-u_k+i)(u_j-u_k+2i)},\\
\theta_3 &= \frac{(u_1-u_2+2i)\,(u_3-u_4+2i)\left[(u_1-u_3+2i)\,(u_2-u_4+3i)+i\,(u_3-u_4)-1\right]}{\prod_{1\leq j< k \leq 4}(u_j-u_k+i)(u_j-u_k+2i)}.\nonumber
\end{align}

Lastly, the matrix part for creating the mixed scalar-fermion state is \cite{Sever:2021nsq}
\begin{align}
\Pi_{0|\psi\phi\phi\bar{\psi}}(0|v_1,u_1,u_2,v_2) = &-\frac{3i}{4}\,\frac{1}{\left(u_1-v_1-\frac{3i}{2}\right)\left(u_2-v_1-\frac{3i}{2}\right)\left(u_1-v_2+\frac{3i}{2}\right)\left(u_2-v_2+\frac{3i}{2}\right)}\nn\\ &\times\frac{(u_1+u_2)\,(v_1+v_2)-2\,u_1u_2-2\,v_1v_2+5i\,(v_1-v_2)-\frac{21}{2}}{(u_1-u_2+i)\,(u_1-u_2+2i)\,(v_1-v_2+3i)}\,.
\end{align}
The most non-trivial four-particle matrix parts that arise are thus for four scalars, two scalars and two fermions, as well as for four fermions.

\subsection{Non-trivial contractions}
Most contractions of the matrix parts of the creation amplitude and the FF transition are trivial. The only noteworthy cases correspond to four fermions and four scalars.

For four fermions, contracting the last line of \eqref{eq: 4pt FFbar and psipsibar} with \eqref{eq: four scalar pentagon patrix part} yields
\begin{equation}
\label{eq: four fermion matrix part contraction}
\begin{aligned}
 & \Pi_{0|\psi\psi\bar{\psi}\bar{\psi}}(0|u_1,u_2,v_1,v_2)^{A_1A_2}_{B_1B_2}F_{\psi^{\text{}}_{A_1}\psi'_{A_2}\bar{\psi}'_{B_1}\bar{\psi}^{\text{}}_{B_2}}(u_1,u_2,v_1,v_2)\\
 &=\frac{4\,(u_1-u_2)^2+22}{\left((u_1-u_2)^2+1\right)\left((u_1-u_2)^2+4\right)}\\&\times\frac{\delta(u_1-v_2)\,\delta(u_2-v_1)-S_{\psi\psi}(u_1,u_2)\,\delta(u_1-v_2)\,\delta(u_2-v_1)}{\mu_\psi(u_1)\mu_\psi(u_2)}
 \,,
 \end{aligned}
\end{equation}
where we used that \cite{Belitsky:2016vyq}
\begin{equation}
 S_{B_1B_2}^{A_1A_2}(u,v)=S_{\psi\psi}(u,v)\left[\delta_{B_1}^{A_1}\delta_{B_2}^{A_2}s^{(1)}_{\psi\psi}(u,v)+\delta_{B_2}^{A_1}\delta_{B_1}^{A_2}s^{(2)}_{\psi\psi}(u,v)\right]\,,
\end{equation}
with
\begin{equation}
s^{(1)}_{\psi\psi} = \frac{u - v}{u - v - i}\,, \quad s^{(2)}_{\psi\psi} = \frac{-\,i}{u - v - i}\,.
\end{equation}
When multiplying \eqref{eq: four fermion matrix part contraction} with the dynamical part of the pentagon transition, the two parts in \eqref{eq: four fermion matrix part contraction} that are related via the S-matrix give identical contributions, such that one could drop the second term in favour of a factor of $2$.

For four scalars, we contract $\Pi_{i_1i_2i_3 i_4}(u_1,u_2,u_3,u_4)$ in \eqref{fourscalars2} with \eqref{eq: four scalar pentagon matrix part}, finding 
\begin{equation}
 \begin{aligned}
 &\Pi_{0|\phi\phi\phi\phi}(0|u_1,u_2,u_3,u_4)_{i_1i_2i_3i_4}\Pi_{i_1i_2i_3 i_4}(u_1,u_2,u_3,u_4)\\
 &=
 \frac{9}{\prod_{1\leq j< k \leq 4}(u_j-u_k+i) (u_j-u_k+2 i) }\sum_{j=1}^4\biggl(
 2 u_j^2 u_{j+1}^2
 -7 u_{j}^2 u_{j+1} u_{j+2}
 +3 u_{j} u_{j+1}^2 u_{j+2}
 \\
 &\qquad-7 u_{j} u_{j+1} u_{j+2}^2
 +\frac{7}{2} u_{j}^2 u_{j+2}^2+4 u_j^2+\frac{11}{2} u_{j} u_{j+1} u_{j+2} u_{j+3}+14 u_{j} u_{j+1}-18 u_{j} u_{j+2}+16\biggr)
 \end{aligned}
\end{equation}
Note that this expression is not real, only becoming real after it is multiplied by the factors $(u_j-u_k-i)(u_j-u_k-2i)$ in $\Fphiphibar(u_j,u_k)$ in the dynamical part.
The other two structures in \eqref{fourscalars}, $\Pi^{\sigma_{34}}_{i_1i_2i_3i_4}(u_1,u_2,u_3,u_4)$ and $\Pi^{\sigma_{23}}_{i_1i_2i_3i_4}(u_1,u_2,u_3,u_4)$, are related to the one above via Watson, and they thus yield identical contributions after including also the dynamical part. Thus, they effectively contribute a factor of $3$.

\section{R-matrices with fermions}\la{fermionsR}
In this appendix, we summarize the matrix form of the S-matrices with fermions that are used in section \ref{ferandglue}. For more details, we refer the reader to appendix A of \cite{Basso:2014koa}.

The S-matrix between two fermions or between a fermion and an anti-fermion factorizes into a coupling-dependent part and the R-matrix%
\footnote{It is a non-trivial fact that the coupling-dependent parts of these two are the same.}
\beq\la{Spsipsi}
S_{\psi\psi}(u,v)_{AB}^{CD}=S_{\psi\psi}(u,v)R_{{\bf 4 4}}(u-v)_{AB}^{CD}\,,\qquad S_{\psi\bar\psi}(u,v)_{AD}^{CB}=S_{\psi\psi}(u,v)R_{{\bf 4}\bar{\bf 4}}(u-v)_{AD}^{CB}\,,
\eeq
where
\beq\la{Rpsipsi}
R_{{\bf 4 4}}(w)_{AB}^{CD}={w\over w-i}\delta_A^C\delta_B^D-{i\over w-i}\delta_A^D\delta_B^C\,,\qquad R_{{\bf 4}\bar{\bf 4}}(w)_{AD}^{CB}=\delta_A^C\delta_D^B-{i\over w-2i}\delta_A^B\delta_D^C\,.
\eeq
The conjugated R-matrices are trivially related to the ones given above, 
\beq\la{Rpsipsiconj}
R_{{ \bar{\bf 4} \bar{\bf 4}}}(w)^{AB}_{CD}= R_{{\bf 4 4}}(w)^{AB}_{CD}\,,\qquad R_{\bar{\bf 4} {\bf 4}}(w)_{CB}^{AD} = R_{{\bf 4}\bar{\bf 4}}(w)_{BC}^{DA}\,,
\eeq

Similarly, for a fermion and a scalar, we have
\beq\label{SandR}
 S_{\phi\psi}(u, v)_{iA}^{jB} =  S_{\phi\psi}(u, v) R_{\textbf{64}}(u-v)_{iA}^{jB}\, , \qquad R_{\textbf{64}}(w)_{iA}^{jB} = \delta_{i}^{j}\delta_{A}^{B} + \frac{i}{2w-3i}\rho_{iAC}\rho^{jCB}\, ,
\eeq
The $\rho$-matrices entering the R-matrices~\eqref{SandR} are the off-diagonal components of the 6D Dirac $\gamma$-matrices in the Weyl representation. They satisfy the Clifford algebra
\beq
\rho_{iAC}\rho^{jCB} + \rho_{jAC}\rho^{iCB} = 2\delta_{i}^{j}\delta_{A}^{B}\, , \qquad \textrm{where} \qquad \rho^{iAB} = -\rho^{iBA} = -\rho^*_{iAB}\, ,
\eeq
and are usually referred to as 't Hooft symbols.

\bibliographystyle{utphys2}
\bibliography{bib}

\end{document}